\documentclass[12pt,a4paper]{article}
	
\usepackage{a4wide}
\usepackage{amsmath}
\usepackage{amssymb}
\usepackage{amsfonts}
\usepackage{epsfig}
\usepackage{exscale}
\usepackage{float}
\usepackage{bbm}
\usepackage[numbers,sort&compress]{natbib}

\newcommand{\R}{{\mathbb{R}}}
\newcommand{\C}{{\mathbb{C}}}

\makeatletter
\@addtoreset{equation}{section}
\makeatother

\title{Asymptotic Freedom, Dimensional Transmutation, \\
and an Infra-red Conformal Fixed Point for the \\
$\delta$-Function Potential in 1-dimensional \\
Relativistic Quantum Mechanics}

\author{M.\ H.\ Al-Hashimi$^{a,b}$, A.\ M.\ Shalaby$^{a,c}$, and 
U.-J.\ Wiese$^{b}$ 
\footnote{Contact information: M.\ H.\ Al-Hashimi: hashimi@itp.unibe.ch,
+41 31 631 8878; A.\ Shalaby, amshalab@qu.edu.qa, +974 4403 4630; 
U.-J.\ Wiese, wiese@itp.unibe.ch, +41 31 613 8504.}
\\ \\
$^a$ Department of Mathematics, Statistics, and Physics \\
Qatar University, Al Tarfa, Doha 2713, Qatar \\ \\
$^b$ Albert Einstein Center for Fundamental Physics \\
Institute for Theoretical Physics, Bern University \\
Sidlerstrasse 5, CH-3012 Bern, Switzerland \\ \\
$^c$ Physics Department, Faculty of Science \\
Mansoura University, Egypt \\ \\}

\begin{document} 

\maketitle

\vspace{-1cm}

\begin{abstract} \normalsize

We consider the Schr\"odinger equation for a relativistic point particle in an
external 1-dimensional $\delta$-function potential. Using dimensional 
regularization, we investigate both bound and scattering states, and we obtain 
results that are consistent with the abstract mathematical theory of 
self-adjoint extensions of the pseudo-differential operator 
$H = \sqrt{p^2 + m^2}$. Interestingly, this relatively simple system 
is asymptotically free. In the massless limit, it undergoes dimensional
transmutation and it possesses an infra-red conformal fixed point. Thus it can 
be used to illustrate non-trivial concepts of quantum field theory in the 
simpler framework of relativistic quantum mechanics.

\end{abstract}

\newpage
 
\section{Introduction}

The unification of quantum physics and special relativity is achieved in the 
framework of relativistic quantum field theories. In particular, in the standard
model of particle physics elementary particles are very successfully described 
as quantized wave excitations of the corresponding quantum fields. As such, they
have qualitatively different properties than the point particles of Newtonian 
mechanics or quantum mechanics. In particular, while the position of a quantum
mechanical point particle is in general uncertain, quantized waves do not even 
have a conceptually well-defined position in space. Unlike in quantum 
mechanics, in local quantum field theory a ``particle'' is a non-local object
\cite{New49,Ree61,Rui81,Fle00,Hal01}. It is 
well-known that a unification of point particle mechanics and special 
relativity is problematical, even at the classical level. In particular, Currie,
Jordan, and Sudarshan proved that two point particles cannot interact in 
such a way that the principles of special relativity are respected, i.e.\ that 
the system provides a representation of the Poincar\'e algebra \cite{Cur63}. 
Leutwyler has generalized this result to an arbitrary number of particles 
\cite{Leu65}. His non-interaction theorem states that classical relativistic 
point particles are necessarily free, as a consequence of Poincar\'e invariance.
The only exception are two particles in one spatial dimension confined to each
other by a linearly rising potential. In one dimension, the corresponding 
confining string has no other degrees of freedom than the positions of its
endpoints, which are represented by the two point particles. While strings can
interact relativistically in higher dimensions, according to Leutwyler's 
non-interaction theorem, point particles can not. Hence, it is not surprising
that particle physics is based on quantum field theory rather than on 
relativistic point particle quantum mechanics. It should also be noted that, by 
including the interaction in the momentum and not in the boost operator, 
interesting relativistic systems with a fixed number of interacting particles 
have been constructed and investigated in detail \cite{Rui80,Rui87,Rui01}.
However, in this case, the coordinates and momenta of the particles do not obey
canonical commutation relations, and thus do not describe ordinary point
particles.

When studying fundamental physics, it is a big step to proceed from 
non-relativistic quantum mechanics to relativistic quantum field theory. Not
only for pedagogical reasons, it is interesting to ask whether non-trivial 
systems of relativistic quantum mechanics exist. Even free quantum mechanical 
relativistic point particles have some interesting properties 
\cite{Bak73,Alm84,Str06,AlH09}.
Minimal position-velocity wave packets of such particles spread in such a way 
that probability leaks out of the light-cone. While such a quantum mechanical
violation of causality does not happen in relativistic quantum field theories, 
it would arise in a hypothetical world of relativistic point particles 
\cite{Fle65,Heg74,Heg80,Heg85,Ros87,Mos90,Heg01,Bar03,Bus05,Bus06}. While in 
quantum field theory a local Hamiltonian gives rise to non-local field 
excitations that manifest themselves as ``particles'', in relativistic quantum 
mechanics local point particles of mass $m$ follow the dynamics of the non-local
Hamiltonian $H = \sqrt{p^2 + m^2}$. According to Leutwyler's non-interaction
theorem, one cannot add a potential to this Hamiltonian without violating the
principles of relativity theory, already at the classical level. This is not
surprising, because a potential would describe instantaneous interactions at a 
distance, mediated with infinite speed. The only exception are singular contact 
interactions, which are not excluded by the classical non-interaction theorem. 
Hence, there might be a quantum loop-hole in the theorem, which would be worth 
exploring, at least for pedagogical reasons, trying to bridge the large gap 
between non-relativistic quantum mechanics and relativistic quantum field 
theory in studying fundamental physics.

In non-relativistic quantum mechanics, contact interactions have been studied
in great detail 
\cite{Cal88,Tho79,Beg85,Hag90,Jac91,Fer91,Gos91,Mea91,Man93,Phi98},
which has been used to illustrate some non-trivial concepts of quantum field 
theories in the simpler context of non-relativistic quantum mechanics.
In this paper we endow the Hamiltonian 
$H = \sqrt{p^2 + m^2}$ for a single free relativistic point particle in one 
spatial dimension with a contact interaction potential $\lambda \delta(x)$. We 
can imagine that such a potential is generated by a second particle of infinite
mass. Once this case is fully understood, as a next step one can then consider
two relativistic particles of finite mass, and ask whether a contact interaction
leads to a non-trivial representation of the Poincar\'e group, thus providing a 
quantum mechanical loop-hole in the classical non-interaction theorem. In this
paper, we do not yet address that question and limit ourselves to a single
particle in the external 1-dimensional $\delta$-function potential. 
Remarkably, already this relatively simple problem provides interesting insights
into some qualitative differences between relativistic and non-relativistic 
quantum mechanics. While the simple $\delta$-function potential provides a 
standard textbook problem, a non-relativistic particle moving in one spatial 
dimension allows more general contact interactions. It can actually distinguish 
a 4-parameter family of such interactions. This follows from the theory of 
self-adjoint extensions \cite{Neu32,Ree75} of the local free-particle kinetic 
energy Hamiltonian $H = \frac{p^2}{2m}$ 
\cite{Ber61,Alb88,Jac91,Car90,Bon01,AlH12}. There is a 4-parameter family of
self-adjoint extensions characterized by the boundary condition for the wave 
function at the contact point
\begin{equation}
\left(\begin{array}{c} \Psi(\varepsilon) \\ \partial_x \Psi(\varepsilon)
\end{array}\right) = 
\exp(i \theta) \left(\begin{array}{cc} a & b \\ c & d \end{array}\right) 
\left(\begin{array}{c} \Psi(-\varepsilon) \\ \partial_x \Psi(-\varepsilon)
\end{array}\right).
\end{equation}
Here $\varepsilon \rightarrow 0$, $a, b, c, d \in \R$ with $ad - bc = 1$, and
$\theta \in ]- \frac{\pi}{2},\frac{\pi}{2}]$. The five parameters 
$a, b, c, d, \theta$ with the constraint $ad - bc = 1$ provide a 
4-parameter family of self-adjoint extensions of the non-relativistic 
free-particle Hamiltonian, and thus a 4-parameter family of quantum mechanical 
contact interactions. The standard contact interaction potential 
$\lambda \delta(x)$ just corresponds to $a = d = 1$, $b = 0$, $c = 2 m \lambda$,
and $\theta = 0$. The most general contact interaction does not respect parity 
symmetry, which requires $a = d$ and $\theta = 0$. Still, in the 
non-relativistic case, this leaves a 2-parameter family of parity-invariant 
contact interactions. A free particle with a generalized energy-momentum 
dispersion relation $H = \sum_{n= 0}^N c_n p^n$ even allows an $N^2$-parameter 
family of self-adjoint extensions. For very high momenta $p$, the energy of such
a particle increases as $p^N$, which for $N > 2$ allows the resolution of 
further details of a contact point than for the standard non-relativistic 
dispersion relation with $N = 2$. If one thinks of the relativistic 
energy-momentum dispersion relation $H = \sqrt{p^2 + m^2}$ as a power series 
expansion in $p^2$
with $N \rightarrow \infty$, in the relativistic case one might perhaps expect 
an infinite number of self-adjoint extension parameters, and thus an infinite
variety of contact interactions, e.g.\ represented by the $\delta$-function
potential and all its derivatives. However, the opposite is true. At large
momentum $p$, the relativistic energy $\sqrt{p^2 + m^2}$ only increases as
$|p|$, which provides less short-distance resolution than the non-relativistic
$p^2$. Indeed, there is just a 1-parameter family of self-adjoint extensions of
the relativistic free-particle Hamiltonian $H = \sqrt{p^2 + m^2}$, which can be
characterized by the parameter $\lambda$ in the contact interaction potential
$\lambda \delta(x)$. This follows from the self-adjoint extension theory of
so-called pseudo-differential operators, which includes the non-local 
Hamiltonian $H = \sqrt{p^2 + m^2}$ \cite{Alb97}. This theory also predicts that 
in higher dimensions, relativistic point particles are completely unaffected by 
contact interactions and thus remain free. This is again in contrast to the 
non-relativistic case, in which there is a 1-parameter family of contact 
interactions both in two and in three spatial dimensions \cite{Jac91}.

As a result of Leutwyler's non-interaction theorem as well as of the theory of
self-adjoint extensions of the pseudo-differential operator 
$H = \sqrt{p^2 + m^2}$, relativistic quantum mechanics is a rather narrow 
subject. In particular, for a single particle one is limited to the simple
$\delta$-function or to a linear confining potential. In this context, it is 
important to point out that
the Klein-Gordon and Dirac equations do not belong to relativistic quantum
mechanics, but to quantum field theory. In particular, it is well-known that
these equations do not allow a consistent single-particle interpretation,
because they address the physics of both particles and anti-particles. The
relativistic point particle Hamiltonian $H = \sqrt{p^2 + m^2}$, on the other
hand, is concerned just with particles. The problem of the relativistic
$\delta$-function potential has already been investigated in the mathematical
literature as an application of the theory of self-adjoint extensions of
pseudo-differential operators \cite{Alb97}. Here we address the problem using
more traditional tools of theoretical physics. Unlike in the non-relativistic
case, the relativistic $\delta$-function potential gives rise to ultra-violet
divergences which we regularize and renormalize using dimensional 
regularization \cite{Bol72,Bol72a,tHo72,Bie14}. It is reassuring that the 
results that we obtain are indeed consistent with those obtained by the 
self-adjoint extension theory of Ref.\cite{Alb97}. Here we study the system 
in great detail, and 
address various interesting physics questions, including strong bound states 
with a binding energy that exceeds the rest mass of the bound particle. 
Remarkably, this relatively simple quantum mechanical model shares several 
non-trivial features with relativistic quantum field theories. In particular, 
just like Quantum Chromodynamics (QCD) \cite{Fri73}, it is asymptotically free 
\cite{Gro73,Pol73}.

In two spatial dimensions, a non-relativistic $\delta$-function potential must 
also be renormalized 
\cite{Cal88,Tho79,Beg85,Hag90,Jac91,Fer91,Gos91,Mea91,Man93,Phi98}.
While this system is classically scale invariant, at the
quantum level it dynamically generates a bound state via dimensional 
transmutation, and it has scattering states which display asymptotic freedom. 
Hence, it can be used to illustrate these non-trivial features, which are 
usually encountered in quantum field theory, in the framework of 
non-relativistic quantum mechanics. However, this theory can not be obtained as 
the non-relativistic limit of a relativistic theory. In this paper, we show 
that asymptotic freedom and 
dimensional transmutation already arise in 1-dimensional relativistic point 
particle quantum mechanics with a $\delta$-function potential. Furthermore, in 
the massless limit the system is scale-invariant, at least at the classical 
level. However, just like in QCD, scale invariance is anomalously 
broken at the quantum level. The system then undergoes dimensional 
transmutation and generates a mass scale non-perturbatively. Unlike QCD, in the 
massless limit the relativistic quantum mechanical model even has a free 
infra-red conformal fixed point. Although actual elementary particles are 
quantized waves rather than point-like objects, addressing these topics in 
relativistic point particle quantum mechanics makes them more easily accessible 
than just studying them in the standard context of relativistic quantum field 
theories.

The rest of this paper is organized as follows. In section 2, we consider the
bound state problem and use the bound state energy to define a renormalization
condition. In section 3, we derive the relativistic probability current density
and show explicitly that it is conserved. In section 4, we address the
scattering states and we show that the energy-dependent running coupling
constant is finite after renormalization. Reflection and transmission 
amplitudes, as well as the scattering phase shift, the scattering length, and
the effective range are derived in section 5. In section 6, we investigate the
energy-dependence of the running coupling constant and its $\beta$-function, and
we show that the theory is asymptotically free. In section 7, we study 
ultra-strong bound states and the corresponding scattering states. Section 8 
analyzes the massless limit, in which the system undergoes dimensional 
transmutation, and develops an infra-red conformal fixed point. Finally, 
section 9 contains our conclusions.

\section{Dimensional Regularization and Renormalization of a Bound State}

Let us consider the relativistic time-independent Schr\"odinger equation 
\begin{equation}
\sqrt{p^2 + m^2} \Psi(x) + \lambda \delta(x) \Psi(x) = E \Psi(x).
\end{equation}
In momentum space
\begin{equation}
\Psi(x) = \frac{1}{2 \pi} \int dp \ \widetilde \Psi(p) \exp(i p x), \quad
\Psi(0) = \frac{1}{2 \pi} \int dp \ \widetilde \Psi(p), \
\end{equation}
and the Schr\"odinger equation takes the form
\begin{equation}
\label{Schroedinger}
\sqrt{p^2 + m^2} \ \widetilde \Psi(p) + 
\frac{\lambda}{2 \pi} \int dp' \ \widetilde \Psi(p') = E \widetilde \Psi(p),
\end{equation}
such that for a bound state
\begin{equation}
\widetilde \Psi_B(p) =
\frac{\lambda \Psi_B(0)}{E_B - \sqrt{p^2 + m^2}}.
\end{equation}
Integrating this equation over all momenta, we obtain the gap equation
\begin{eqnarray}
&&\Psi_B(0) = \frac{1}{2 \pi} \int dp \ \widetilde \Psi_B(p) =
\lambda \Psi_B(0) \frac{1}{2 \pi} \int dp \ \frac{1}{E_B - \sqrt{p^2 + m^2}} \
\Rightarrow \nonumber \\
&&\frac{1}{\lambda} = 
\frac{1}{2 \pi} \int dp \ \frac{1}{E_B - \sqrt{p^2 + m^2}},
\end{eqnarray}
which determines the bound state energy $E_B$.
The resulting integral is logarithmically ultra-violet divergent and must hence 
be regularized. We do this by using dimensional regularization, i.e.\ by 
analytically continuing the spatial dimension to $D = 1 + \varepsilon \in \C$ 
and by finally taking the limit $\varepsilon \rightarrow 0$. While the coupling 
constant $\lambda$ is dimensionless in one dimension, in $D$ dimensions the 
prefactor of the $\delta$-function has dimension (mass)$^{1-D}$. In order to 
renormalize the bare coupling, we let it depend on the cut-off, and we replace 
$\lambda$ by $\lambda(\varepsilon) m^{-\varepsilon}$. In order to keep 
$\lambda(\varepsilon)$ dimensionless, we have factored out the dimensionful
term $m^{-\epsilon} = m^{1-D}$, using the particle mass $m$ as the renormalization
scale. The regularized gap equation then takes the form
\begin{equation}
\label{lambda}
\frac{m^{D-1}}{\lambda(\varepsilon)} = 
\frac{1}{(2 \pi)^D} \int d^Dp \ \frac{1}{E_B - \sqrt{p^2 + m^2}} = I(E_B).
\end{equation}
For a bound state $E_B < m$, and we expand the integrand in powers of 
$E_B/\sqrt{p^2 + m^2}$, such that
\begin{equation}
I(E_B) = - \frac{\pi^{D/2} D}{\Gamma(D/2 + 1)} \int_0^\infty dp \ 
\frac{p^{D-1}}{\sqrt{p^2 + m^2}} 
\sum_{n=0}^\infty \left(\frac{E_B}{\sqrt{p^2 + m^2}}\right)^n.
\end{equation}
While all higher-order terms are finite, the leading term (with $n = 0$) is 
logarithmically ultra-violet divergent. All terms can be integrated separately,
and then be re-summed, which in the limit $\varepsilon \rightarrow 0$ yields
\begin{equation}
\label{IEB}
I(E_B) = m^\varepsilon \left[\frac{1}{\pi \varepsilon} +
\frac{\gamma - \log(4 \pi)}{2 \pi} - \frac{E_B}{2 \pi \sqrt{m^2 - E_B^2}} 
\left(\pi + 2 \arcsin\frac{E_B}{m}\right)\right].
\end{equation}
Here $\gamma \approx 0.5772$ is Euler's constant. For an ultra-strong bound 
state with energy $E_B < - m$ the series from above diverges. Still, the result 
can be obtained by directly integrating the convergent expression
\begin{eqnarray}
\label{IEBU}
&&\frac{1}{2 \pi} \int dp \ \left(\frac{1}{E_B - \sqrt{p^2 + m^2}} +
\frac{1}{\sqrt{p^2 + m^2}} \right) = \nonumber \\
&&\frac{E_B}{\pi \sqrt{E_B^2 - m^2}} 
\text{arctanh} \frac{\sqrt{E_B^2 - m^2}}{E_B}.
\end{eqnarray}

As a renormalization condition, we now hold the binding energy $E_B$ fixed in
units of the mass $m$, such that the running bare coupling is given by
\begin{equation}
\frac{1}{\lambda(\varepsilon)} = \frac{1}{\pi \varepsilon} +
\frac{\gamma - \log(4 \pi)}{2 \pi} - \frac{E_B}{2 \pi \sqrt{m^2 - E_B^2}} 
\left(\pi + 2 \arcsin\frac{E_B}{m}\right).
\end{equation}

Let us consider the non-relativistic limit, in which the binding energy
$\Delta E_B = E_B - m$ is small compared to the rest mass. In that case, the 
running bare coupling is given by
\begin{eqnarray}
\frac{1}{\lambda(\varepsilon)}&=&\frac{1}{\pi \varepsilon} +
\frac{\gamma - \log(4 \pi)}{2 \pi} - \frac{E_B}{2 \pi \sqrt{m^2 - E_B^2}} 
\left(\pi + 2 \arcsin\frac{E_B}{m}\right) \nonumber \\
&\rightarrow&\frac{1}{\pi \varepsilon} + \frac{\gamma - \log(4 \pi)}{2 \pi} -
\sqrt{\frac{m}{- 2 \Delta E_B}}.
\end{eqnarray}
Interestingly, for the non-relativistic contact interaction $\lambda \delta(x)$,
which does not require renormalization, for $\lambda < 0$ the bound state 
energy is given by $\Delta E_B = - m \lambda^2/2$ such that
$1/\lambda = - \sqrt{- m/2 \Delta E_B}$. This suggests to define a  
renormalized coupling
\begin{equation}
\label{lambdaEB}
\frac{1}{\lambda(E_B)} = - \frac{E_B}{2 \pi \sqrt{m^2 - E_B^2}} 
\left(\pi + 2 \arcsin\frac{E_B}{m}\right) < 0,
\end{equation}
which is defined at the scale $E_B$. Dropping the terms
$1/\pi \varepsilon + [\gamma - \log(4 \pi)]/2 \pi$ corresponds to the modified 
minimal subtraction $\overline{MS}$ scheme that is commonly used in quantum 
field theory.

Let us now determine the bound state wave function in coordinate space
\begin{equation}
\Psi_B(x) = \frac{A}{2 \pi} \int dp \ 
\frac{\exp(i p x)}{E_B - \sqrt{p^2 + m^2}}.
\end{equation}
The integration can be extended to the closed contour $\Gamma$ illustrated in
Figure 1. 
\begin{figure}[t]
\begin{center}
\epsfig{file=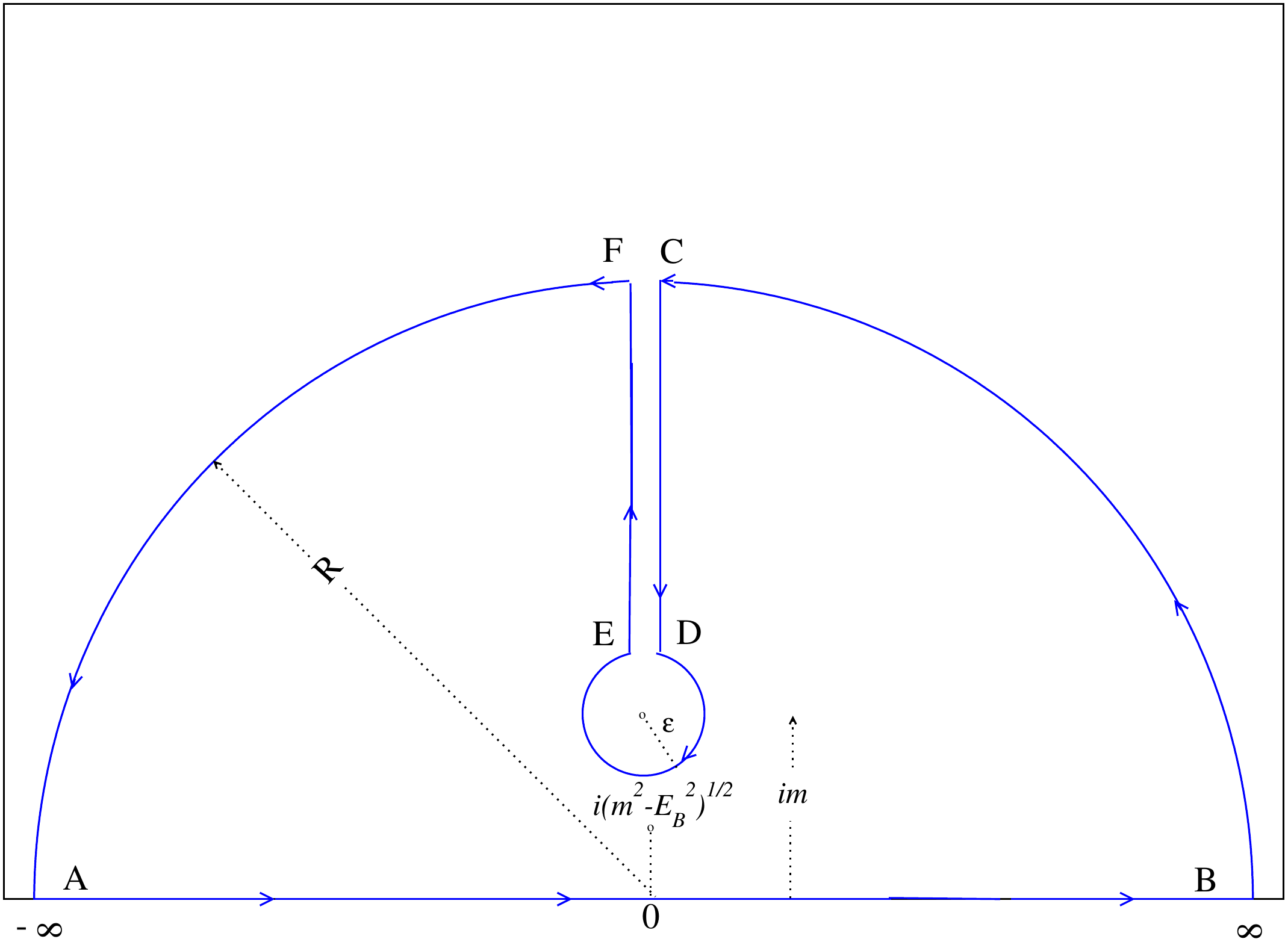,width=0.74\textwidth} \vskip0.5cm
\epsfig{file=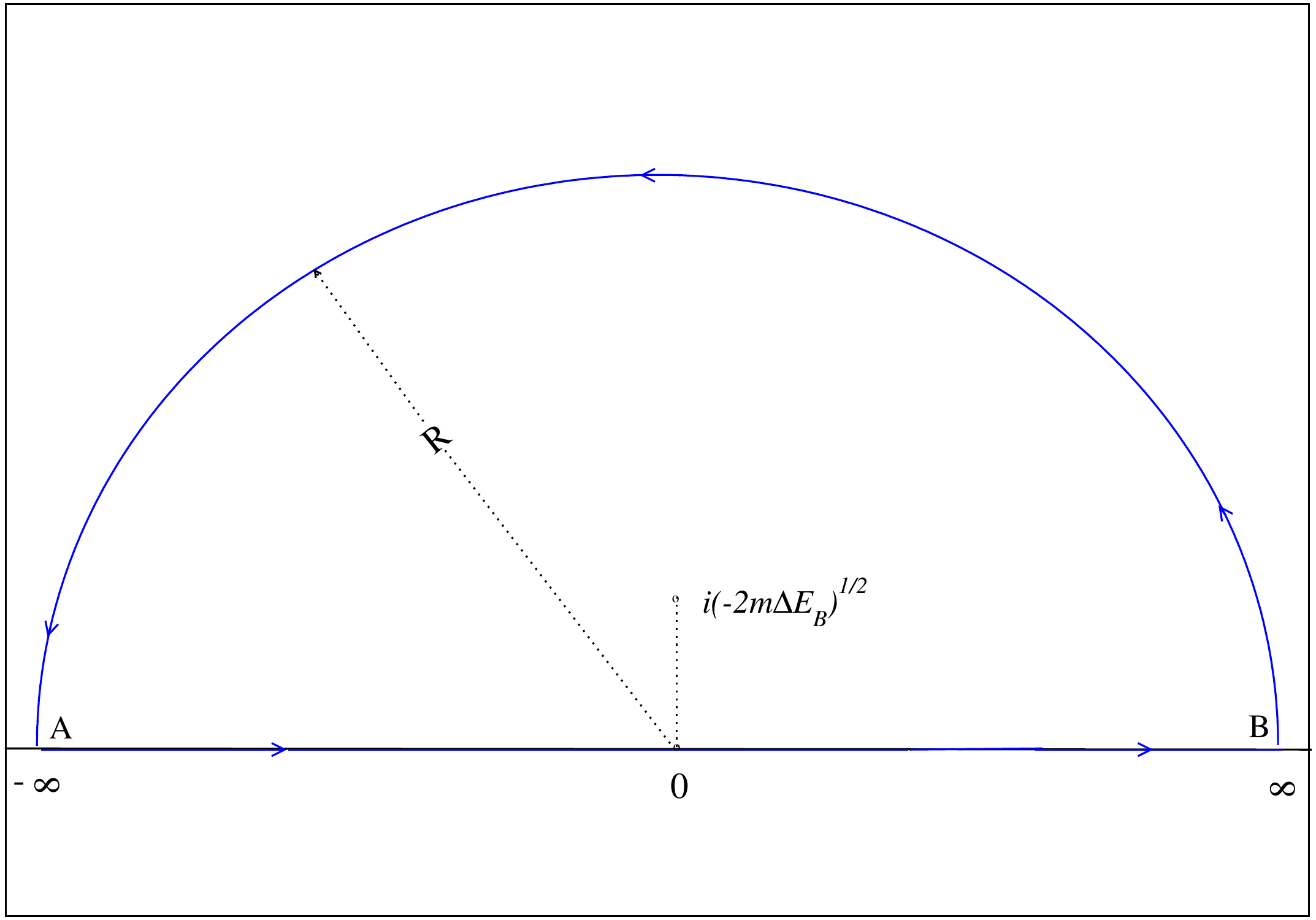,width=0.74\textwidth}
\end{center}
\caption{\it Integration contours for the determination of the wave function of 
the bound state. In the relativistic case (top panel), there is a branch cut 
along the positive imaginary axis, starting at $p = im$. In addition, for 
$0 < E_B < m$, there is a pole at $p = i \sqrt{m^2 - E_B^2}$. In the 
non-relativistic case (bottom panel), there is still a pole, but no branch cut.}
\end{figure}
For $0 < E_B < m$, the integrand has a pole at 
$p = i \sqrt{m^2 - E_B^2}$, which is enclosed by $\Gamma$, as well as a branch 
cut along the positive imaginary axis starting at $p = i m$. The wave function 
then takes the form
\begin{equation}
\label{boundstate}
\Psi_B(x) = A \left[\frac{1}{\pi}  \int_m^\infty d\mu 
\frac{\sqrt{\mu^2 - m^2}}{E_B^2 - m^2 + \mu^2} \exp(- \mu |x|) +
\frac{E_B \exp(- \sqrt{m^2 - E_B^2} |x|)}{\sqrt{m^2 - E_B^2}}\right].
\end{equation}
The integral results from the two contributions along the branch cut, while the
last term is the residue of the pole at $p = i \sqrt{m^2 - E_B^2}$. As 
illustrated in Fig.2, the wave function is logarithmically divergent at the 
origin. This short-distance divergence is unaffected by the renormalization. In 
particular, the singularity of the wave function is integrable and it is thus 
normalizable in the usual sense. 
\begin{figure}[t]
\begin{center}
\epsfig{file=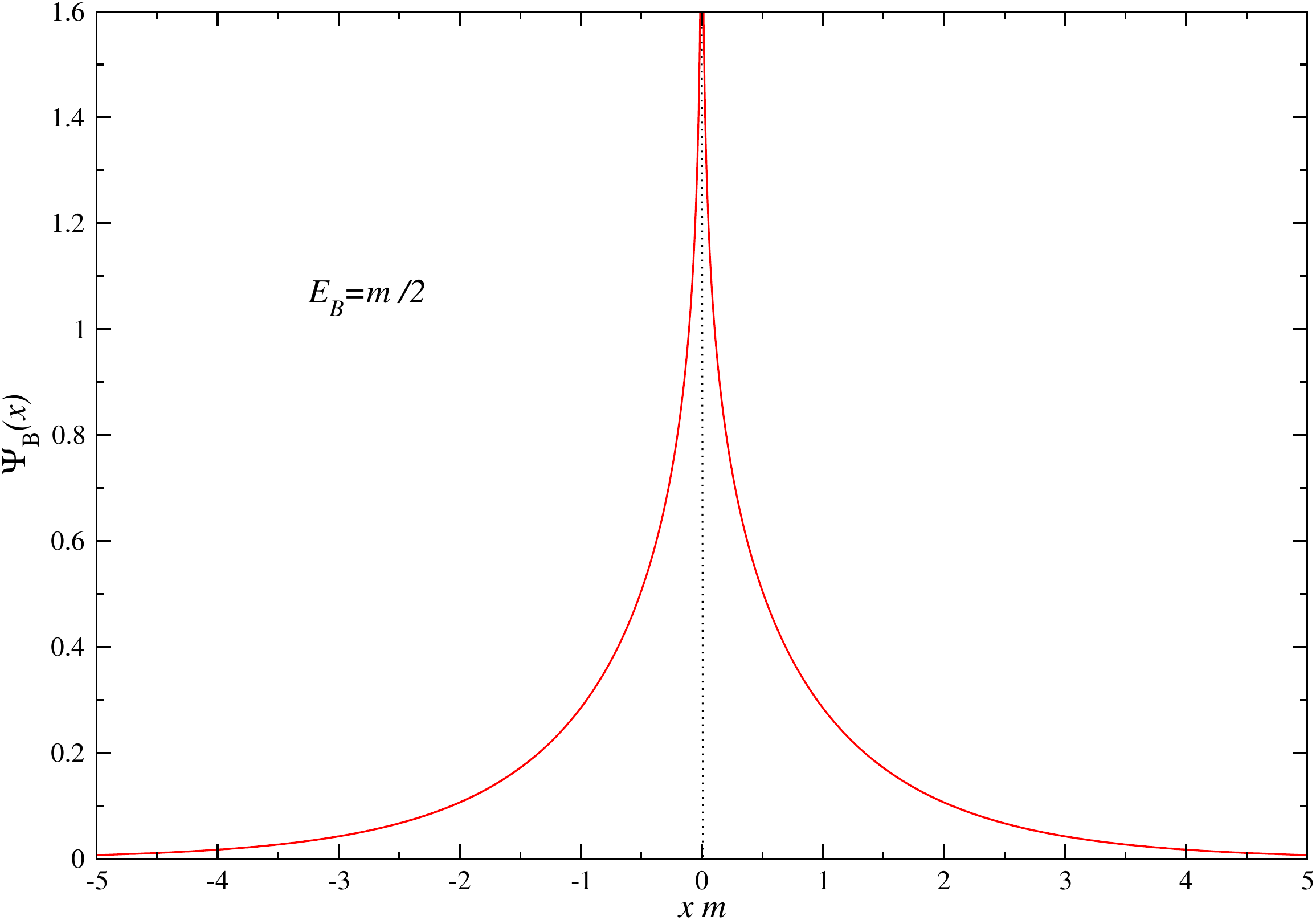,width=0.65\textwidth} \vskip0.5cm
\epsfig{file=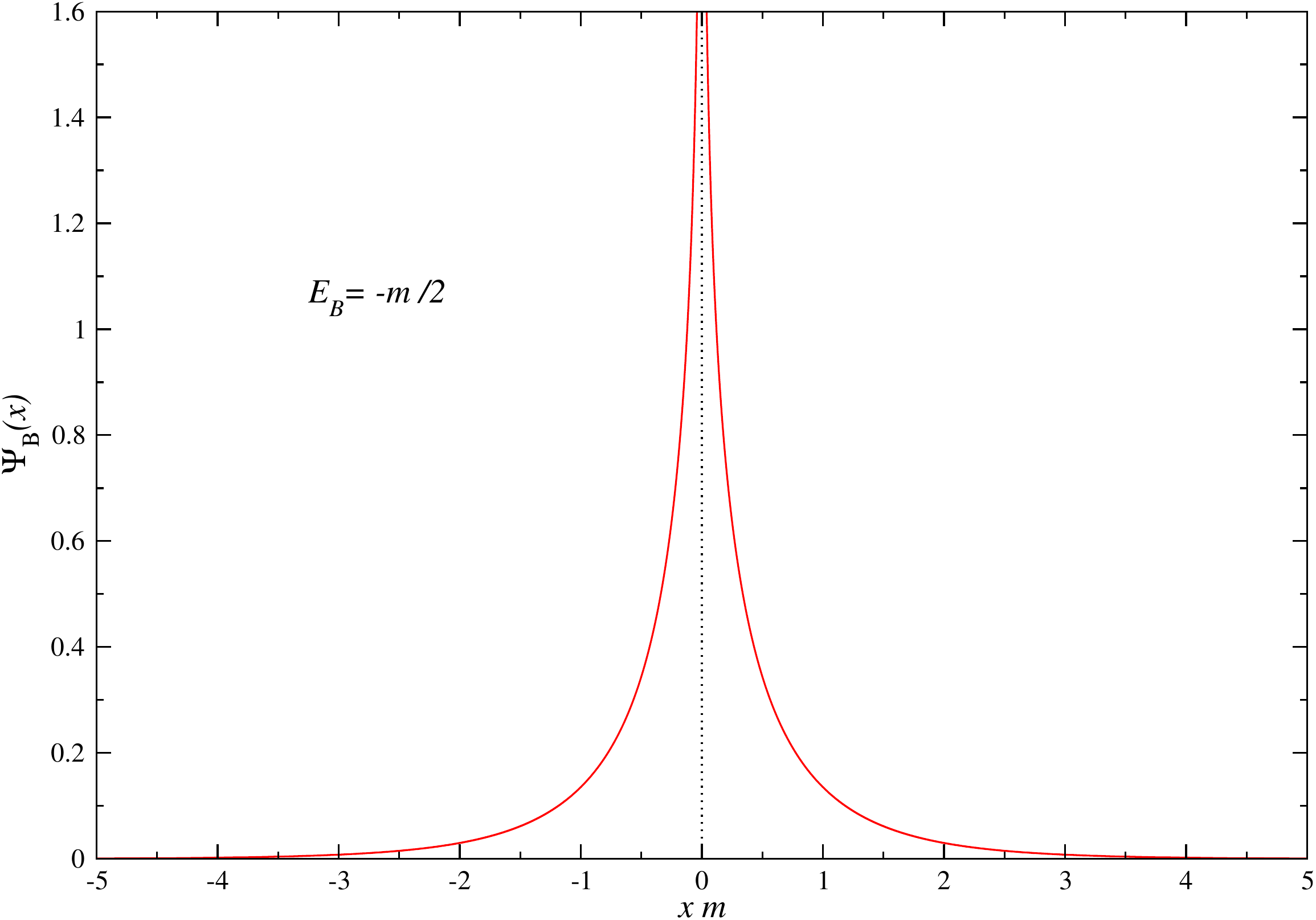,width=0.65\textwidth}
\end{center}
\caption{\it Bound state wave function in coordinate space for an ordinary
bound state with $E_b = m/2$ (top panel), and for a strong bound state with
$E_b = - m/2$ (bottom panel).}
\end{figure}
Alternatively, the bound state wave function can be expressed in terms
of Bessel functions
\begin{equation}
\Psi_B(x) = \frac{A}{\sqrt{\pi}} \sum_{n=0}^\infty \left(\frac{E_B}{m}\right)^n
\left(\frac{m |x|}{2}\right)^{n/2} 
\frac{K_{n/2}(m |x|)}{\Gamma\left(\frac{n+1}{2}\right)}.
\end{equation}
The normalization constant is most easily determined in momentum space
\begin{eqnarray}
&&\frac{|A|^2}{2 \pi} \int dp \frac{1}{(E_B - \sqrt{p^2 + m^2})^2} = 1 \
\Rightarrow \nonumber \\
&&\frac{2 \pi}{|A|^2} = \frac{2 E_B}{m^2 - E_B^2} + 
\frac{m^2}{(m^2 - E_B^2)^{3/2}}\left(\pi + 2 \arcsin\frac{E_B}{m}\right).
\end{eqnarray}
For the non-relativistic $\delta$-function potential, the wave
function is finite at the origin and given by
\begin{equation}
\label{nonrelbound}
\Psi_B(x) = \sqrt{\varkappa} \exp(- \varkappa |x|), \quad 
\Delta E_B = - \frac{\varkappa^2}{2m}. 
\end{equation}
In the non-relativistic limit, the relativistic wave function of 
eq.(\ref{boundstate}) reduces to
\begin{equation}
\Psi_B(x) = \sqrt{\varkappa}
\left[\frac{\varkappa}{m \pi}  \int_m^\infty d\mu 
\frac{\sqrt{\mu^2 - m^2}}{\mu^2 - \varkappa^2} \exp(- \mu |x|) +
\exp(- \varkappa |x|)\right].
\end{equation}
Since $\varkappa/m \rightarrow 0$ in the non-relativistic limit, it indeed
reduces to the non-relativistic wave function of eq.(\ref{nonrelbound}). 
However, the divergence of the relativistic wave function persists for any 
non-zero value of $\varkappa/m$. As we
discussed in the introduction, a non-relativistic contact interaction is 
characterized by a 4-parameter family of self-adjoint extensions, while in
the relativistic case there is only a 1-parameter family (parametrized by 
$\lambda$). The other non-relativistic contact interactions can not be obtained
by taking the non-relativistic limit of a relativistic theory.

Finally, let us also consider the strong bound states, for which the bound 
state energy $E_B < 0$, i.e.\ the binding energy $\Delta E_B = E_B - m$ even 
exceeds the rest mass. In that case, the ``pole'' at $p = i \sqrt{m^2 - E_B^2}$ 
has a vanishing residue and hence does not contribute to the result. The wave 
function of a strong bound state then takes the form
\begin{equation}
\Psi_B(x) = \frac{A}{\pi}  \int_m^\infty d\mu
\frac{\sqrt{\mu^2 - m^2}}{E_B^2 - m^2 + \mu^2} \exp(- \mu |x|), \quad E_B < 0.
\end{equation}
The wave functions for a bound state with $E_B = m/2$ and for a strong
bound state with energy $E_B = - m/2$ are illustrated in Figure 2. One would
think that a relativistic system should not have negative total energy. In 
fact, the total energy should at least be as large as the positive rest mass of 
the system. In our case, translation invariance is explicitly broken by the 
contact potential, which means that the previous argument is not applicable 
here. One may think of the contact interaction as being generated by an 
infinitely heavy second particle located at $x = 0$. When the infinite mass of
this particle is included in the total energy, it is indeed positive.

\section{The Relativistic Probability Current}

In the non-relativistic Schr\"odinger equation, probability is conserved because
of the continuity equation
\begin{equation}
\label{continuity}
\partial_t \rho(x,t) + \partial_x j(x,t) = 0,
\end{equation}
which relates the probability density $\rho(x,t) = |\Psi(x,t)|^2$ to the 
probability current density $j(x,t) = \frac{1}{2 m i} 
[\Psi(x,t)^* \partial_x \Psi(x,t) - \partial_x \Psi(x,t)^* \Psi(x,t)]$.
While in the Dirac and Klein-Gordon equations, probability conservation is 
violated due to the presence of anti-particles, in the relativistic
Schr\"odinger equation discussed here, there are no anti-particles and the
continuity equation (\ref{continuity}) still holds with the usual probability
density $\rho(x,t) = |\Psi(x,t)|^2$, however, with the modified relativistic
probability current density, whose leading terms are
\begin{eqnarray}
j(x,t)&=&\frac{1}{2 m i} 
[\Psi(x,t)^* \partial_x \Psi(x,t) - \partial_x \Psi(x,t)^* \Psi(x,t)] 
\nonumber \\
&+&\frac{1}{8 m^3 i} 
[\Psi(x,t)^* \partial^3_x \Psi(x,t) - 
\partial_x \Psi(x,t)^* \partial_x^2 \Psi(x,t) \nonumber \\
&+&\partial^2_x\Psi(x,t)^* \partial_x \Psi(x,t) - 
\partial^3_x \Psi(x,t)^* \Psi(x,t)] + \dots
\end{eqnarray}
In momentum space the divergence $\partial_x j(x,t)$ takes the compact form
\begin{equation}
p \widetilde j(p,t) = \frac{1}{2 \pi} \int dq \ \widetilde \Psi(-q,t)^* 
[\sqrt{(p-q)^2 + m^2} - \sqrt{q^2 + m^2}] \widetilde \Psi(p-q,t).
\end{equation}
This expression trivially generalizes to an arbitrary energy-momentum dispersion
relation $E(p)$ and yields
\begin{equation}
\widetilde j(p,t) = \frac{1}{2 \pi} \int dq \ \widetilde \Psi(-q,t)^* 
\frac{1}{p}[E(p-q) - E(q)] \widetilde \Psi(p-q,t).
\end{equation}
For a general dispersion relation, the bound state wave function in momentum
space takes the form
\begin{equation}
\widetilde \Psi_B(p) = \frac{A}{E_B - E(p)}.
\end{equation}
The divergence of the probability density then automatically vanishes because
\begin{eqnarray}
p \widetilde j(p)&=&\frac{|A|^2}{2 \pi} \int dq \ \frac{1}{E_B - E(q)}
[E(p-q) - E(q)] \frac{1}{E_B - E(p-q)} \nonumber \\
&=&\frac{|A|^2}{2 \pi} \int dq
\left(\frac{1}{E_B - E(p-q)} - \frac{1}{E_B - E(q)}\right) = 0.
\end{eqnarray}

\section{Dimensional Regularization and Renormalization of Scattering States}

Let us now consider the scattering states. First of all, the states of odd 
parity, which vanish at the origin, are unaffected by the $\delta$-function
potential. Hence, we limit ourselves to stationary scattering states of even 
parity, which we parametrize as 
\begin{equation}
\widetilde \Psi_E(p) = \delta(p - \sqrt{E^2 - m^2}) + 
\delta(p + \sqrt{E^2 - m^2}) + \widetilde \Phi_E(p).
\end{equation}
Later we will combine scattering states of even and odd parity in order to 
extract the reflection and transmission amplitudes. Inserting the ansatz from
above in eq.(\ref{Schroedinger}), we obtain
\begin{eqnarray}
\label{scatteringstate}
&&(\sqrt{p^2 + m^2} - E) \widetilde \Phi_E(p) + \frac{\lambda}{\pi} +
\frac{\lambda}{2 \pi} \int dp' \ \widetilde \Phi_E(p') = 0 \ \Rightarrow 
\nonumber \\
&& \widetilde \Phi_E(p) = 
\frac{\lambda}{\pi} \frac{1 + \pi \Phi_E(0)}{E - \sqrt{p^2 + m^2}}.
\end{eqnarray}
Integrating eq.(\ref{scatteringstate}) over all momenta one finds
\begin{equation}
\Phi_E(0) = \frac{1}{2 \pi} \int dp \ \widetilde \Phi_E(p) = \frac{\lambda}{\pi}
[1 + \pi \Phi_E(0)] \frac{1}{2 \pi} \int dp \ \frac{1}{E - \sqrt{p^2 + m^2}}.
\end{equation}
Again, by replacing $\lambda$ with $\lambda(\varepsilon) m^{-\varepsilon}$, and by
using dimensional regularization, we then obtain
\begin{equation}
\Phi_E(0) = \frac{1}{\pi} \frac{\lambda(\varepsilon) m^{-\varepsilon} I(E)}
{1 - \lambda(\varepsilon) m^{-\varepsilon} I(E)}.
\end{equation}
For positive energy $E$ the integral takes the form
\begin{equation}
\label{IE}
I(E) = m^\varepsilon \left[\frac{1}{\pi \varepsilon} +
\frac{\gamma - \log(4 \pi)}{2 \pi} + \frac{E}{\pi \sqrt{E^2 - m^2}} 
\text{arctanh}\frac{\sqrt{E^2 - m^2}}{E}\right].
\end{equation}
Using eq.(\ref{lambda}), the function $\widetilde \Phi_E(p)$ then results as
\begin{equation}
\widetilde \Phi_E(p) = \frac{\lambda(E,E_B)}{\pi} 
\frac{1}{E - \sqrt{p^2 + m^2}},
\end{equation}
with the energy-dependent running coupling constant (again renormalized at the 
scale $E_B$) given by
\begin{eqnarray}
\label{lambdarun}
\lambda(E,E_B) = \frac{1}{I(E_B) - I(E)}&=& 
- \left[\frac{E}{\pi \sqrt{E^2 - m^2}} 
\text{arctanh}\frac{\sqrt{E^2 - m^2}}{E} \right. \nonumber \\
&+&\left. \frac{E_B}{2 \pi \sqrt{m^2 - E_B^2}} 
\left(\pi + 2 \arcsin\frac{E_B}{m}\right)\right]^{-1}.
\end{eqnarray}
Remarkably, using eqs.(\ref{IEB}) and (\ref{IE}), the ultra-violet divergences 
of $I(E)$ and $I(E_B)$ cancel, such that the running coupling constant is 
finite when we take the limit $\varepsilon \rightarrow 0$.

In order to investigate whether the resulting system is self-adjoint, let us now
check the orthogonality of the various states. First, we calculate the scalar
product of the bound state and the scattering states
\begin{eqnarray}
\label{OSB}
\langle \Psi_B|\Psi_E\rangle&=&
\frac{1}{2 \pi} \int dp \ \frac{1}{E_B - \sqrt{p^2 + m^2}} \nonumber \\
&\times&[\delta(p - \sqrt{E^2 - m^2}) + \delta(p + \sqrt{E^2 - m^2}) + 
\widetilde \Phi_E(p)] \nonumber \\
&=&\frac{1}{\pi(E_B - E)} \nonumber \\
&+&\frac{1}{\pi(I(E_B) - I(E))} \frac{1}{2 \pi} 
\int dp \ \frac{1}{E_B - \sqrt{p^2 + m^2}} \ \frac{1}{E - \sqrt{p^2 + m^2}}. 
\nonumber \\ \
\end{eqnarray}
The integral results in
\begin{eqnarray}
&&\frac{1}{2 \pi} 
\int dp \ \frac{1}{E_B - \sqrt{p^2 + m^2}} \ \frac{1}{E - \sqrt{p^2 + m^2}} =
\nonumber \\
&&\frac{1}{E - E_B} \frac{1}{2 \pi} \int dp \ 
\left(\frac{1}{E_B - \sqrt{p^2 + m^2}} - \frac{1}{E - \sqrt{p^2 + m^2}}\right) =
\frac{I(E_B) - I(E)}{E - E_B}, \nonumber \\ \
\end{eqnarray}
such that indeed $\langle \Psi_B|\Psi_E\rangle = 0$. This is also the case for 
a strong bound state with $E_B < 0$, and even for an ultra-strong bound state
with $E_B < - m$. Next we investigate the orthogonality of the scattering states
\begin{eqnarray}
\label{OSS}
\langle \Psi_{E'}|\Psi_E\rangle&=&
\frac{1}{\pi} \delta(\sqrt{E^2 - m^2} - \sqrt{{E'}^2 - m^2}) +
\frac{\lambda(E,E_B)}{\pi^2(E - E')} + \frac{\lambda(E',E_B)}{\pi^2(E' - E)} 
\nonumber \\
&+&\frac{\lambda(E,E_B) \lambda(E',E_B)}{\pi^2} \ \frac{1}{2 \pi} \int dp \ 
\frac{1}{E - \sqrt{p^2 + m^2}} \ \frac{1}{E' - \sqrt{p^2 + m^2}}
\nonumber \\
&=&\frac{1}{\pi} \delta(k - k') + \frac{1}{\pi^2(E - E')}
\left[\frac{1}{I(E_B) - I(E)} - \frac{1}{I(E_B) - I(E')}\right] \nonumber \\
&+&\frac{1}{\pi^2} \ \frac{1}{I(E_B) - I(E)} \ \frac{1}{I(E_B) - I(E')} \ 
\frac{I(E') - I(E)}{E - E'} \nonumber \\
&=&\frac{1}{\pi} \delta(k - k').
\end{eqnarray}
Here we have introduced $k = \sqrt{E^2 - m^2}$ and $k' = \sqrt{{E'}^2 - m^2}$.
The orthogonality of the various states shows explicitly that, after 
regularization and renormalization, the resulting Hamiltonian is indeed 
self-adjoint.

The contour for the determination of the scattering wave function in coordinate 
space is illustrated in Figure 3. In addition to the branch cut, there are two
poles on the real axis at $p = \pm k = \pm \sqrt{E^2 - m^2}$, which give rise to
in- and out-going plane waves. When these poles are avoided by the contour, one 
obtains the contribution $\widetilde \Phi_E(x)$ to the total scattering wave 
function.
\begin{figure}[t]
\begin{center}
\epsfig{file=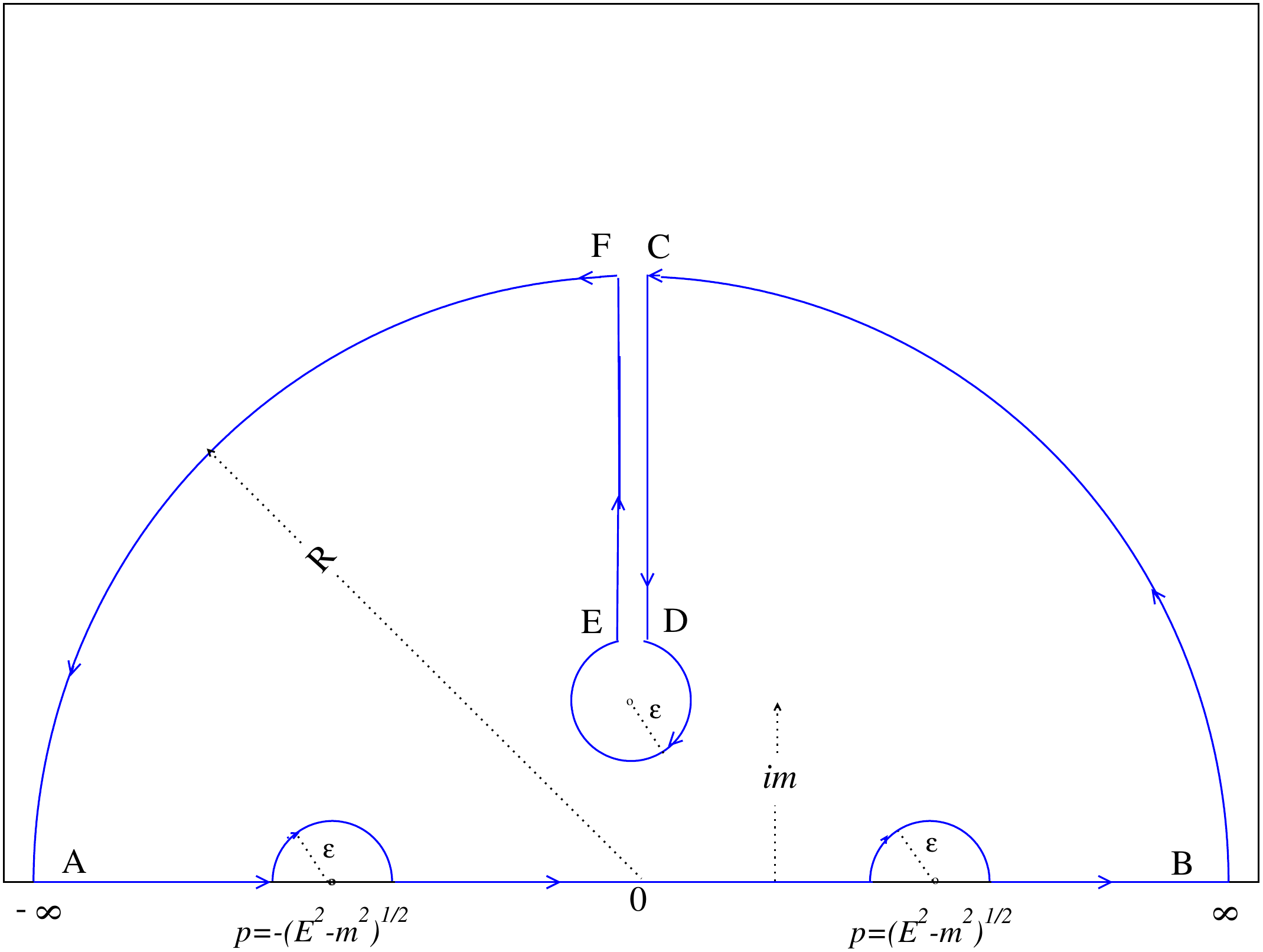,width=0.65\textwidth}
\end{center}
\caption{\it Integration contour for the determination of the wave function of 
the scattering states with $E > m$. There is a branch cut along the positive 
imaginary axis, starting at $p = im$. In addition, there are two poles on the 
real axis at $p = \pm \sqrt{E^2 - m^2}$.}
\end{figure}
The even-parity stationary scattering wave function in coordinate space 
\begin{eqnarray}
\Psi_E(x)&=&A(k) \left[\cos(k x) + 
\lambda(E,E_B) \frac{\sqrt{k^2 + m^2}}{k} \sin(k |x|) \right.
\nonumber \\
&-&\left. \frac{\lambda(E,E_B)}{\pi} \int_m^\infty d\mu 
\frac{\sqrt{\mu^2 - m^2}}{\mu^2 + k^2} \exp(- \mu |x|)\right], \ 
E = \sqrt{k^2 + m^2},
\end{eqnarray}
is illustrated in Figure 4. Like the bound state wave function, it is 
logarithmically divergent at the origin.
\begin{figure}[t]
\begin{center}
\epsfig{file=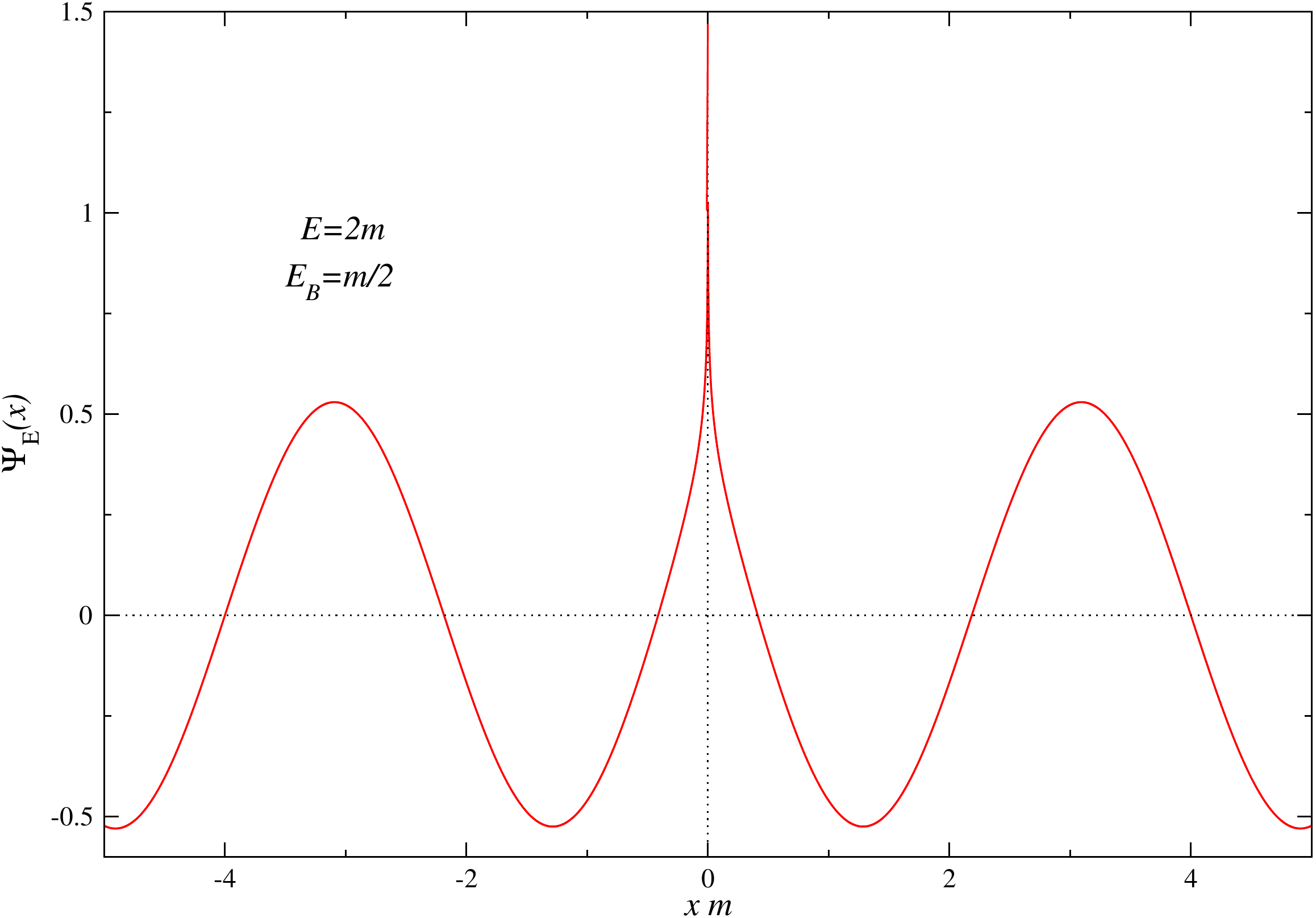,width=0.65\textwidth}
\end{center}
\caption{\it Even-parity stationary scattering wave function in coordinate 
space for $E = 2m$ and $E_B = m/2$. Like the wave function of the bound state, 
the scattering wave function also diverges logarithmically at the origin.}
\end{figure}

Let us again consider the non-relativistic limit by considering small scattering
energies, such that $\Delta E = E - m \ll m$, while also maintaining a small
bound state energy $|\Delta E_B| = |E_B - m| \ll m$. In this case, the 
running coupling constant reduces to
\begin{equation}
\lambda(E,E_B) \rightarrow - \sqrt{- \frac{2 \Delta E_B}{m}} = \lambda,
\end{equation}
where $\lambda$ is indeed the energy-independent coupling constant of the
non-relativistic theory. As for the bound state wave function, the branch-cut 
contribution vanishes in the non-relativistic limit, such that one recovers the
non-relativistic even-parity scattering wave function
\begin{equation}
\Psi_E(x) = A(k) \left[\cos(k x) + \frac{\lambda m}{k} \sin(k |x|)\right].
\end{equation}
Here $k = \sqrt{2 m \Delta E}$. It should be noted that the logarithmic 
divergence at the origin still persists for all non-zero values of 
$\Delta E = k^2/2m$.

\section{Reflection and Transmission Amplitudes}

Let us now construct reflection and transmission amplitudes by superimposing
the non-trivial even-parity scattering states $\Psi_E(x)$ with the trivial 
odd-parity scattering states $B \sin(k x)$. We will now adjust the amplitudes 
$A(k)$ and $B$ of the even and odd scattering states such that the wave function
 takes the form
\begin{equation}
\Psi_I(x) = \exp(i k x) + R(k) \exp(- i k x) + C(k) \lambda(E,E_B) \chi_E(x),
\end{equation}
in region I to the left of the contact point, i.e.\ for $x < 0$. In region II,
for $x > 0$, on the other hand, we demand
\begin{equation}
\Psi_{II}(x) = T(k) \exp(i k x) + C(k) \lambda(E,E_B) \chi_E(x).
\end{equation}
Here
\begin{equation}
\chi_E(x) = \frac{1}{\pi} \int_m^\infty d\mu \
\frac{\sqrt{\mu^2 - m^2}}{\mu^2 + E^2 - m^2} \exp(- \mu |x|),
\end{equation}
is the branch-cut contribution, which arises only in the relativistic case.
Away from the contact point $x = 0$, this contribution decays exponentially and
thus has no effect on the scattering wave function at asymptotic distances.
After a straightforward calculation one obtains
\begin{equation}
A(k) = - C(k) = \frac{k}{k + i \sqrt{k^2 + m^2} \lambda(E,E_B)}, \ B = i,
\end{equation}
which leads to the reflection and transmission amplitudes
\begin{equation}
R(k) = - \frac{i \sqrt{k^2 + m^2} \lambda(E,E_B)}
{k + i \sqrt{k^2 + m^2} \lambda(E,E_B)}, \quad
T(k) = \frac{k}{k + i \sqrt{k^2 + m^2} \lambda(E,E_B)},
\end{equation}
which obey $1 + R(k) = T(k)$. Using eq.(\ref{lambdarun}), it is straightforward 
to convince oneself that $R(k)$ and $T(k)$ have a pole at 
$k = i \sqrt{m^2 - E_B^2}$, which corresponds to the bound state with energy 
$\sqrt{k^2 + m^2} = E_B$. The S-matrix is given by
\begin{equation}
S(k) = R(k) + T(k) =  \frac{k - i \sqrt{k^2 + m^2} \lambda(E,E_B)}
{k + i \sqrt{k^2 + m^2} \lambda(E,E_B)} =
\exp(2 i \delta(k)),
\end{equation}
which determines the scattering phase shift
\begin{equation}
\tan\delta(k) = - \frac{\sqrt{k^2 + m^2} \lambda(E,E_B)}{k}, \quad 
E = \sqrt{k^2 + m^2}.
\end{equation}
In \cite{Alb97}, the problem has been investigated using the self-adjoint 
extension theory of the pseudo-differential operator $\sqrt{p^2 + m^2}$, which 
led to the same expression for the S-matrix. This shows that dimensional 
regularization yields results that are consistent with the more abstract 
mathematical approach. We go significantly beyond the results of 
Ref.\cite{Alb97} by addressing numerous additional physics questions. 

In three dimensions, it is common to consider the low-energy effective range 
expansion, which corresponds to 
$k \cot\widetilde{\delta}(k) = - 1/a_0 + \frac{1}{2} r_0 k^2$, 
where $a_0$ is the scattering length and $r_0$ is the effective range. The
3-d scattering phase shift $\widetilde\delta(k)$ measures the phase of the
outgoing scattering wave relative to a sine-wave that vanishes at the origin.
In our 1-d problem, there is no scattering in the odd-parity sine-wave channel.
The non-trivial 1-d scattering phase $\delta(k)$ measures the phase of the
outgoing scattering wave relative to a cosine-wave that has a maximum at the 
origin. Hence, compared to the 3-d case, $\delta(k)$ corresponds to
$\widetilde{\delta}(k) + \frac{\pi}{2}$, such that $\cot\widetilde{\delta}(k)$
corresponds to $- \tan\delta(k)$. Hence, in our 1-d case, the effective range 
expansion takes the form
\begin{equation}
\label{effectiverange}
- k \tan\delta(k) = - \frac{1}{a} + \frac{1}{2} r_0 k^2 + \dots.
\end{equation}
This yields the scattering length $a_0$ and the effective range $r_0$ as
\begin{equation}
\label{scatteringlength}
a_0 = \frac{1}{m}\left(\frac{1}{\pi} - \frac{1}{\lambda(E_B)}\right), \quad
r_0 = - \frac{1}{a_0 m^2} + \frac{2}{3 \pi a_0^2 m^3}.
\end{equation}
Here $\lambda(E_B) < 0$ is the renormalized coupling constant defined in 
eq.(\ref{lambdaEB}). When there is a bound state, the scattering length is 
positive, and it diverges when the bound state approaches zero energy. In the
absence of a bound state, the scattering lengths would become negative.
The scale of $r_0$ is set by the Compton wave length $1/m$, while its 
particular value is also influenced by the scattering length through the 
dimensionless combination $a m$. The effective range vanishes in the 
non-relativistic limit $a m \rightarrow \infty$, as one might naively expect 
for a contact interaction, but is non-zero in the relativistic case. This is
due to the non-locality of the Hamiltonian $\sqrt{p^2 + m^2}$, which senses the
contact interaction already from some distance $r_0$. The phase shift 
$\delta(k)$ is illustrated in Figure 5. It varies between 
$\delta(0) = \frac{\pi}{2}$ and $\delta(\infty) = 0$. This is consistent with
the 1-d version of Levinson's theorem, which identifies the number of bound 
states as $n = 2 [\delta(0) - \delta(\infty)]/\pi$ \cite{War63,deB94}.
\begin{figure}[t]
\begin{center}
\epsfig{file=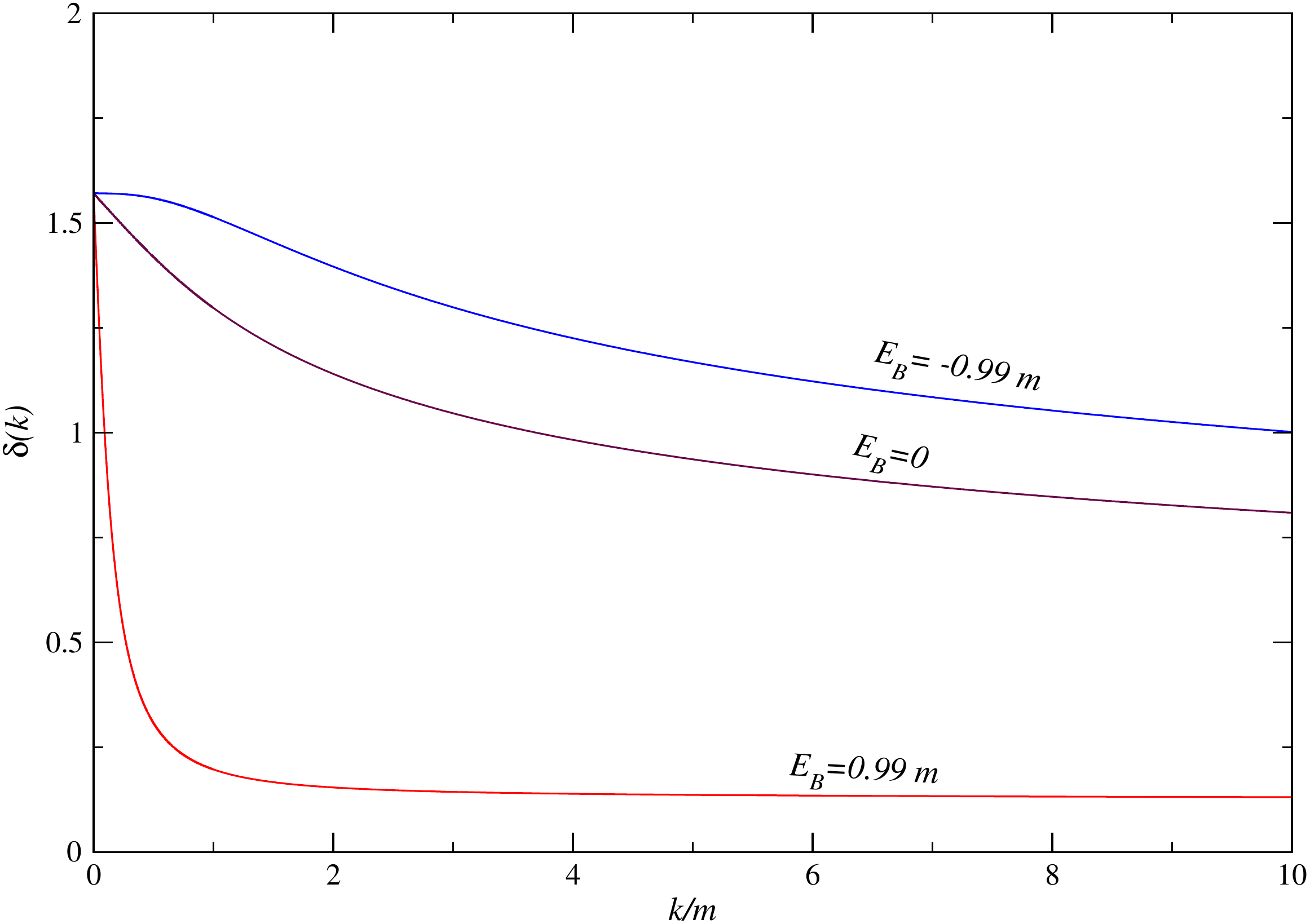,width=0.65\textwidth}
\end{center}
\caption{\it Phase shift $\delta(k)$ as a function of the wave number $k$,
for three different values of $E_B/m = 0.99$, $0$, and $- 0.99$.}
\end{figure}

In the non-relativistic limit, $\lambda(E,E_B)$ again reduces to the 
energy-independent coupling $\lambda$ of the non-relativistic theory, such that
we indeed recover the non-relativistic textbook results
\begin{equation}
R(k) = - \frac{i m \lambda}{k + i m \lambda}, \quad 
T(k) = \frac{k}{k + i m \lambda}, \quad 
S(k) = \frac{k - i m \lambda}{k + i m \lambda}.
\end{equation}
These quantities have a pole at $k = - i m \lambda$, which determines the
non-relativistic bound state energy $\Delta E_B = k^2/2m = - m \lambda^2/2$.
The scattering phase shift $\delta(k)$ is then given by
\begin{equation}
\tan\delta(k) = - \frac{m \lambda}{k},
\end{equation}
which yields the scattering length $a_0 = - 1/(m \lambda)$ and the effective 
range $r_0 = 0$.

\section{Running Coupling Constant, $\beta$-Function, and Asymptotic Freedom}

Until now, we have introduced the coupling $\lambda(E_B)$ of 
eq.(\ref{lambdaEB}), which is renormalized at the bound state energy, as well
as the energy-dependent running coupling $\lambda(E,E_B)$ of 
eq.(\ref{lambdarun}), which again uses $E_B$ as the renormalization 
condition, and enters the reflection and transmission amplitudes in the same way
as the energy-independent coupling $\lambda$ in the non-relativistic case. Let 
us now investigate the dependence of the running coupling $\lambda(E,E_B)$ on 
the scattering energy $E$, which is illustrated in Figure 6.
\begin{figure}[t]
\begin{center}
\epsfig{file=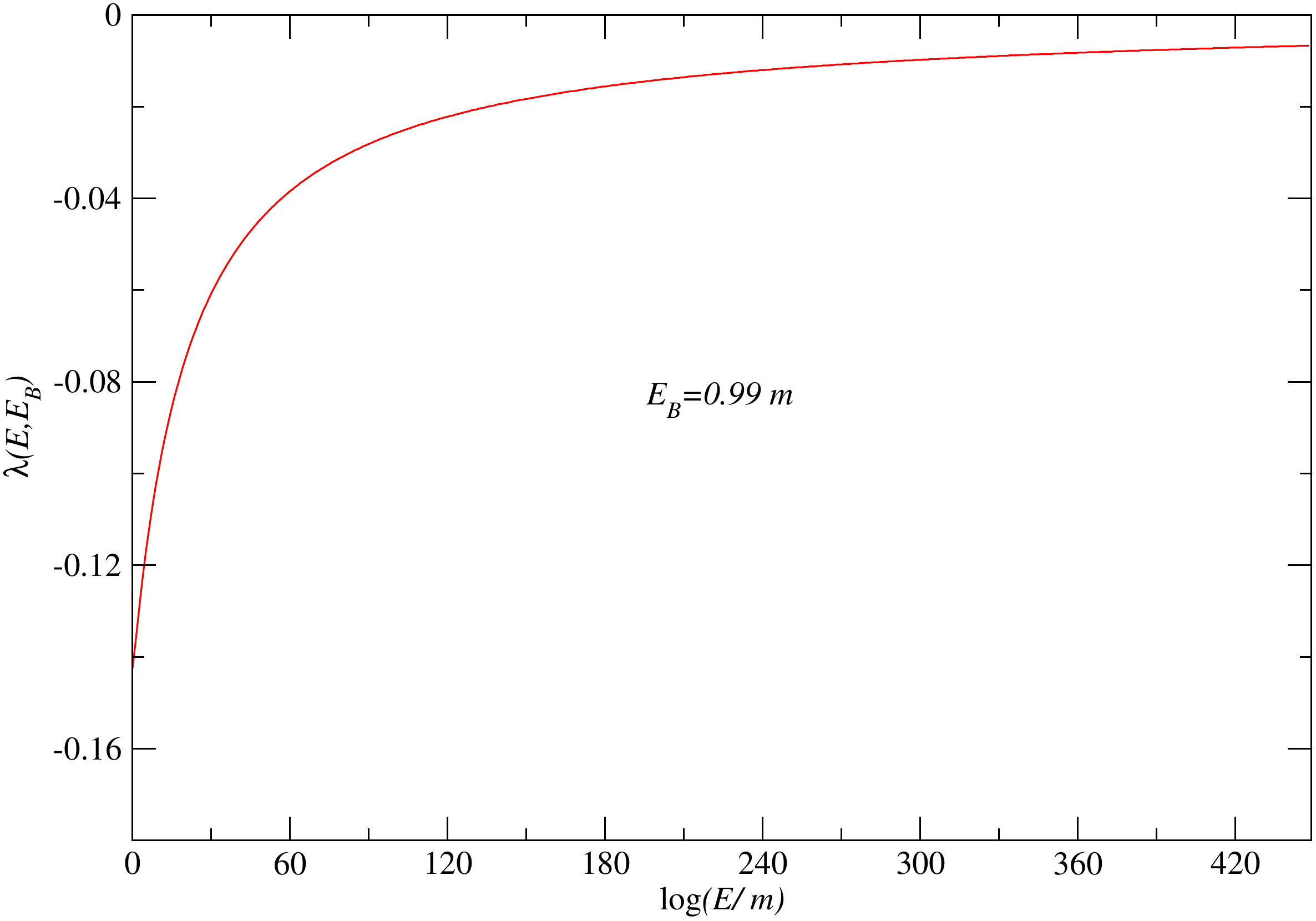,width=0.65\textwidth}
\end{center}
\caption{\it Running coupling $\lambda(E,E_B)$ as a function of the scattering
energy $E$, for $E_B = 0.99 m$.}
\end{figure}

At high energies, $\lambda(E,E_B)$ vanishes logarithmically, thus indicating 
that the scattered particle becomes free in the infinite energy limit. In 
particle physics, e.g.\ in QCD, this behavior is known as asymptotic freedom.
The exact non-perturbative expression for the $\beta$-function takes the form
\begin{eqnarray}
\beta(\lambda(E,E_B))&=&E\frac{\partial |\lambda(E,E_B)|}{\partial E} 
\nonumber \\
&=&- \frac{\lambda(E,E_B)^2}{\pi} +
\frac{\lambda(E,E_B)^2 \epsilon^2}{1 - \epsilon^2} 
\left(\frac{1}{\lambda(E_B)} - \frac{1}{\lambda(E,E_B)} - \frac{1}{\pi}\right). 
\end{eqnarray}
Since $\lambda(E,E_B)$ itself is negative, it is natural to use 
$|\lambda(E,E_B)|$ to define the $\beta$-function. In the above expression, 
$\epsilon = m/E \sim 2 \exp(- \pi/|\lambda(E,E_B)|)$ is non-perturbative
and exponentially suppressed for small $\lambda(E,E_B)$. This implies 
that, to all orders in perturbation theory, the $\beta$-function is given by its
1-loop expression $- \lambda(E,E_B)^2/\pi$. The factor $1/\pi$ plays the role of
the 1-loop coefficient $\beta_0$. Non-perturbative corrections enter through 
$\epsilon$, and become noticeable only at low energies. For asymptotically large
energies, the $\beta$-function behaves as
\begin{equation}
\beta(\lambda(E,E_B)) \rightarrow 
- \frac{\pi}{(\log(E/m))^2} \rightarrow - \frac{\lambda(E,E_B)^2}{\pi} < 0.
\end{equation}
It vanishes at $\lambda(E,E_B) \rightarrow 0$, which corresponds to an 
ultra-violet fixed point. The negative sign of the $\beta$-function again 
signals asymptotic freedom. In Figure 7, the $\beta$-function is illustrated 
for different values of $E_B/m$, which influences the behavior only far away 
from the ultra-violet fixed point at $\lambda(E,E_B) = 0$.
\begin{figure}[t]
\begin{center}
\epsfig{file=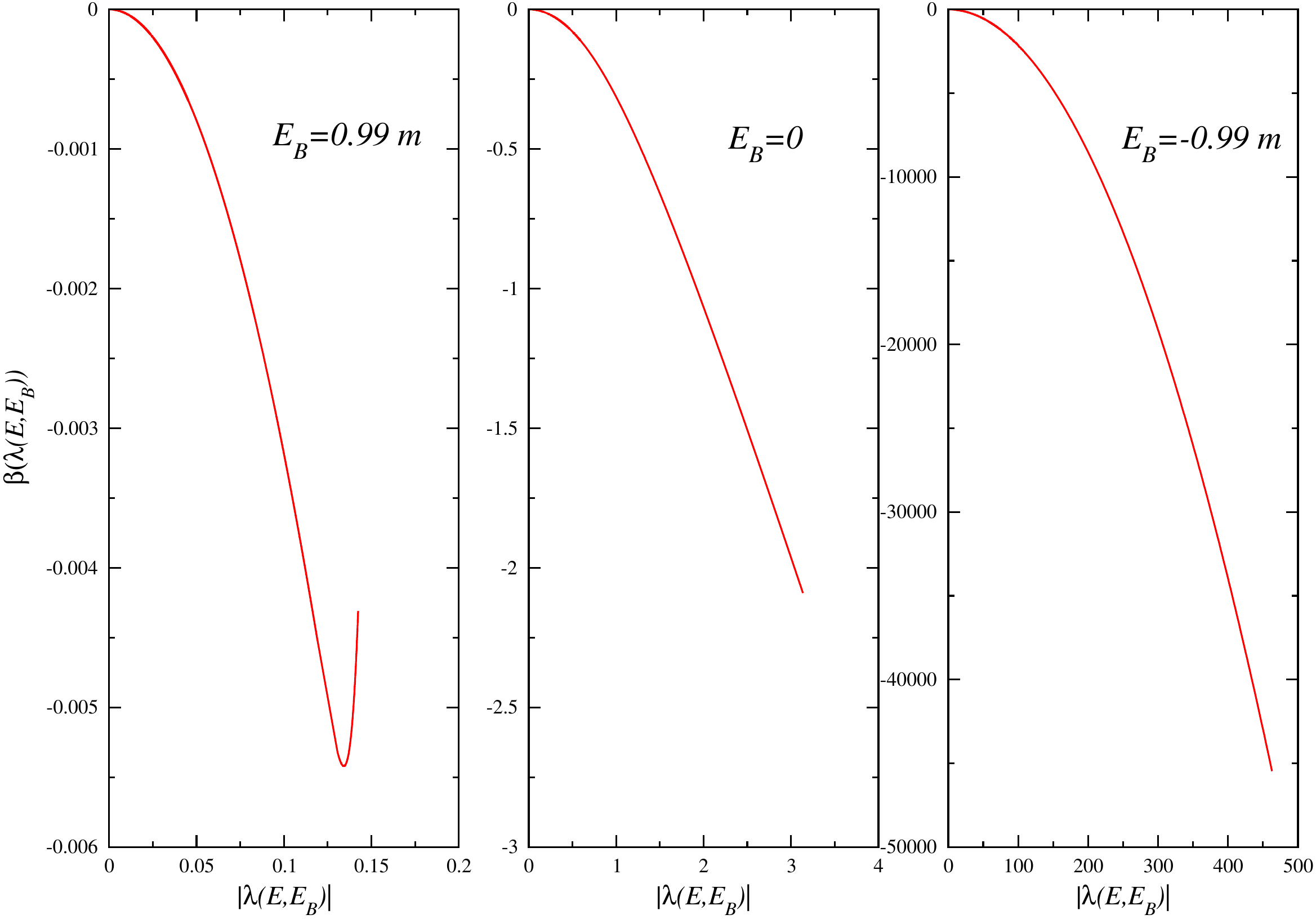,width=\textwidth}
\end{center}
\caption{\it The $\beta$-function $\beta(\lambda(E,E_B))$ as a function of
the running coupling $|\lambda(E,E_B)|$ for three values of $E_B/m = 0.99$,
$0$, and $-0.99$. The end points of the curves correspond to the maximal value
of $|\lambda(m,E_B)|$, which is assumed in the low-energy limit 
$E \rightarrow m$. Note the different scales on the axes, which result from the
very different ranges over which $\lambda(E,E_B)$ is varying.}
\end{figure}

Another zero of the $\beta$-function would require
\begin{equation}
\epsilon^2 \left(\frac{1}{\lambda(E_B)} - \frac{1}{\lambda(E,E_B)}\right) = 
\frac{1}{\pi} \ \Rightarrow \
\frac{E}{m} = \frac{m}{\sqrt{E^2 - m^2}} \text{arctanh} 
\frac{\sqrt{E^2 - m^2}}{E},
\end{equation}
provided that $\epsilon = m/E \neq 1$. However, the above condition is satified 
only for $E = m$, and hence, in this case, no other fixed point exists. As we 
will see in Section 8, in the massless case, $m = 0$, there is an additional 
infra-red conformal fixed point.
 
\section{Ultra-Strong Bound States and Repulsive Scattering States}

Until now, we have used the expression of eq.(\ref{IEB}) for $I(E_B)$, and thus 
we have implicitly assumed that $|E_B| < m$. This includes the case of strong
bound states with $- m < E_B < 0$, but it excludes ultra-strong bound states 
with energies $E_B < - m$. We again point out that the strong and ultra-strong 
bound states are not necessarily tachyonic, because the $\delta$-function 
potential can be attributed to a hypothetical infinitely heavy particle. Hence, 
the total rest energy of the system always remains positive. For $E_B < - m$ one
must use the expression of eq.(\ref{IEBU}) for $I(E_B)$, with
interesting consequences for the bound and scattering state wave functions.
First of all, it should be pointed out that the various states are still
mutually orthogonal, such that the Hamiltonian remains self-adjoint, even in the
presence of an ultra-strong bound state. This is easy to see, because the
orthogonality relations eq.(\ref{OSB}) and eq.(\ref{OSS}) do not depend on the
explicit form of $I(E_B)$.

Let us first consider the extreme limit $E_B \rightarrow - \infty$. In this 
case, the running coupling constant takes the form
\begin{equation}
\lambda(E,E_B) \rightarrow - \left[\frac{E}{\pi \sqrt{E^2 - m^2}} \ 
\text{arctanh}\frac{\sqrt{E^2 - m^2}}{E} - 
\frac{1}{\pi} \log\left(\frac{- 2 E_B}{m}\right)\right]^{-1}.
\end{equation}
For small non-relativistic energies $\Delta E = E - m \ll m$, this reduces to
\begin{equation}
\lambda \rightarrow \frac{\pi}{\log(- 2 E_B/m)} > 0.
\end{equation}
Remarkably, this $\lambda$ actually plays the role of the strength of the 
repulsive contact interaction $\lambda \delta(x)$ in the non-relativistic 
theory. In other words, despite the fact that there is an infinitely strongly
bound state, the low-energy scattering states approach those of the 
non-relativistic repulsive potential $\lambda \delta(x)$, for which there is no 
bound state at all. In fact, in the limit $E_B \rightarrow - \infty$, the 
probability density of the relativistic ultra-strong bound state degenerates to
a $\delta$-function. Because the scattering states still are logarithmically
divergent at the origin, this is not in contradiction with orthogonality in the
non-relativistic limit. Figure 8 compares the even-parity scattering wave
functions at low energy in the relativistic and non-relativistic case, which
indeed coincide, except in the ultimate vicinity of the contact point. This
indeed makes sense, because the short-distance behavior of the relativistic and 
the non-relativistic theory are fundamentally different. We conclude that the 
relativistic contact interaction always produces a bound state. Remarkably, 
when this bound state becomes ultra-strong (with $E_B \rightarrow - \infty$), 
it decouples from the scattering states, which behave as if the contact 
interaction was repulsive. 
\begin{figure}[t]
\begin{center}
\epsfig{file=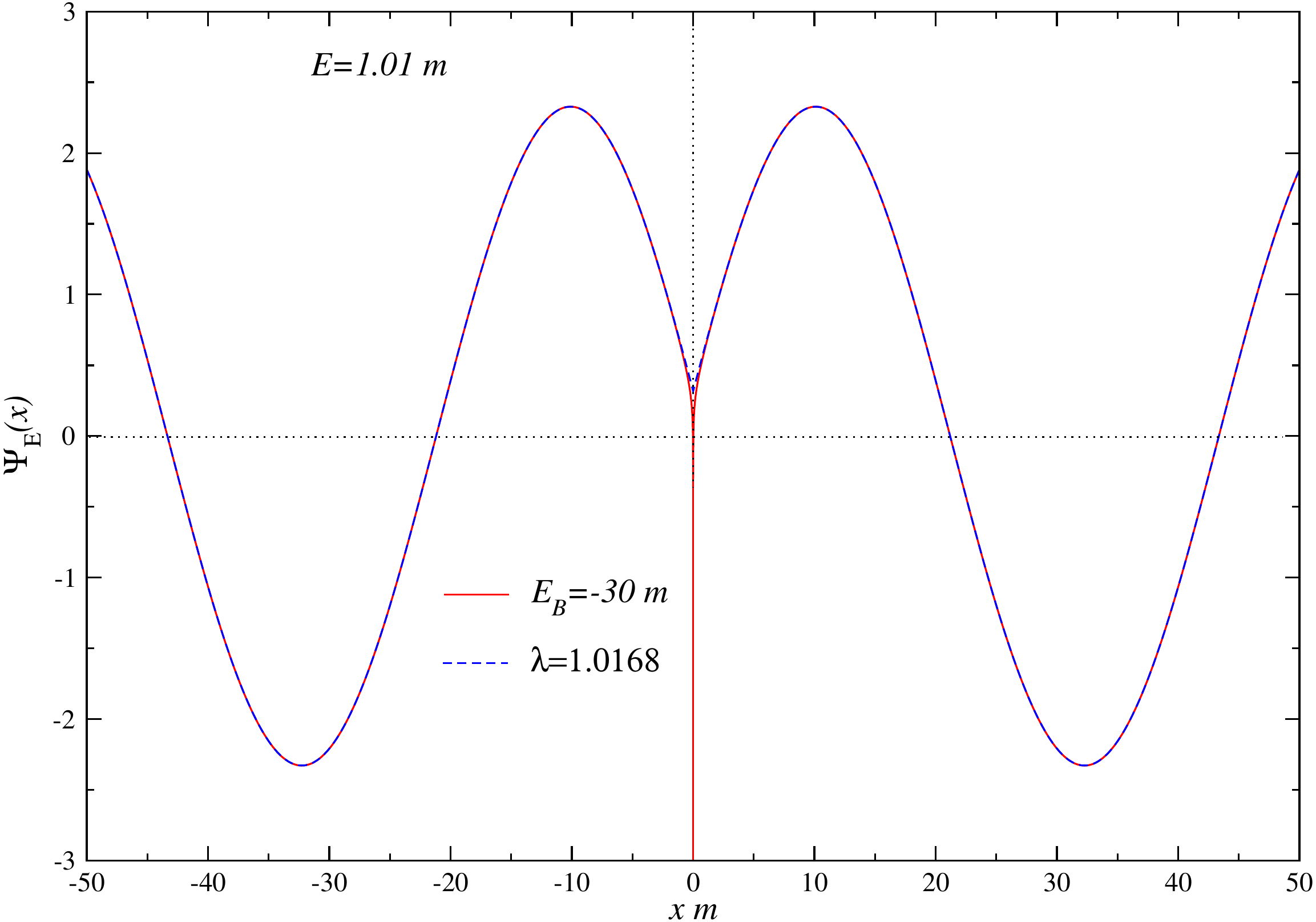,width=0.65\textwidth} \vskip0.5cm
\epsfig{file=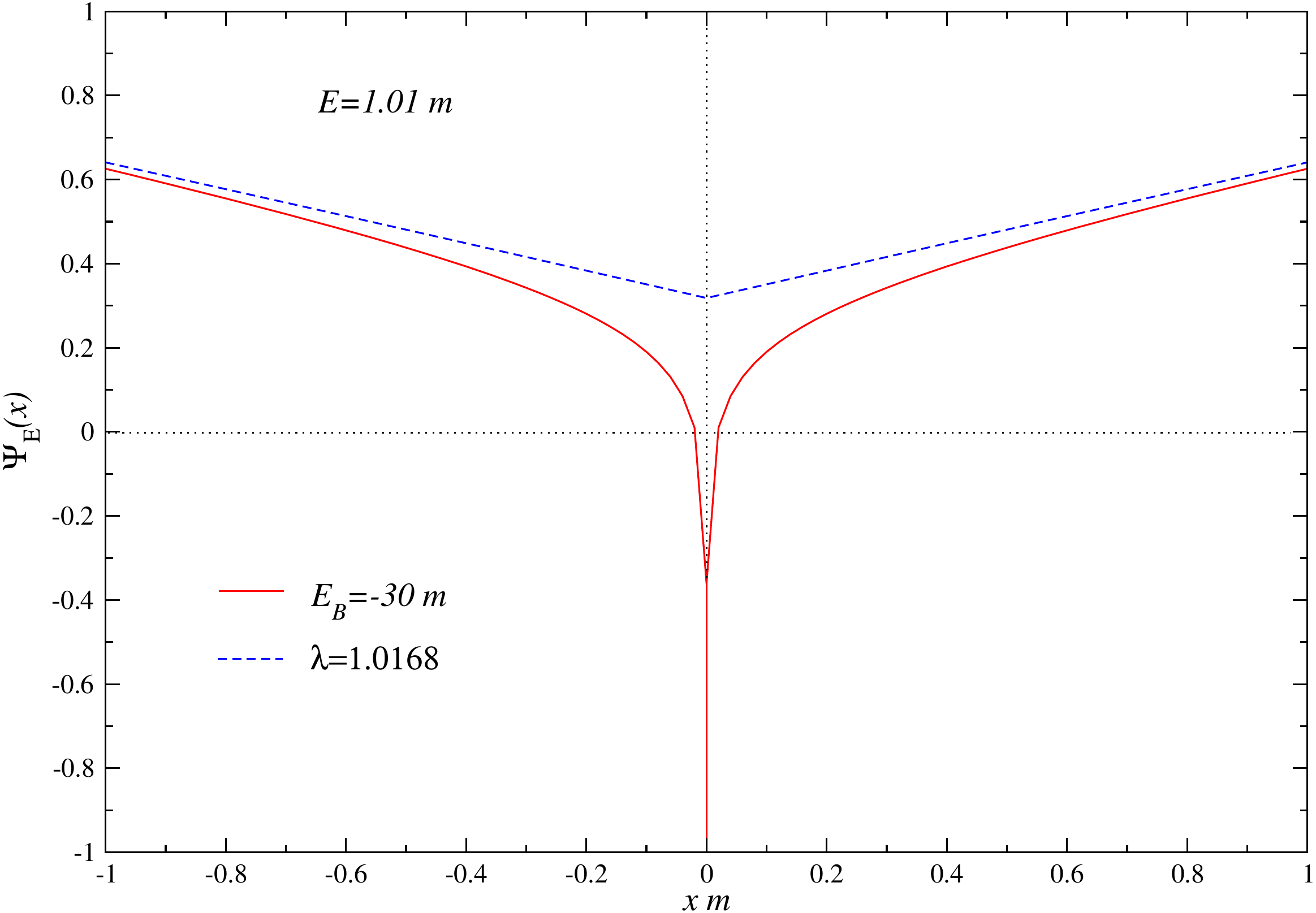,width=0.65\textwidth}
\end{center}
\caption{\it Even-parity low-energy scattering wave function (with energy
$E = 1.01 m$) in the presence of an ultra-strong bound state with 
$E_B = - 30 m$, compared to the corresponding non-relativistic wave function
(with $\lambda = 1.0168$). The panel on the bottom zooms into the region around 
the contact point $x = 0$, in which the relativistic wave function is 
logarithmically divergent, while the non-relativistic wave function remains 
finite.}
\end{figure}

Let us now discuss the case $- \infty < E_B < - m$. In this case, the running
coupling constant is given by
\begin{eqnarray}
\lambda(E,E_B)&=&- \left[\frac{E}{\pi \sqrt{E^2 - m^2}} \ 
\text{arctanh}\frac{\sqrt{E^2 - m^2}}{E} \right. \nonumber \\
&-&\left. \frac{E_B}{\pi \sqrt{E_B^2 - m^2}} 
\text{arctanh} \frac{\sqrt{E_B^2 - m^2}}{E_B}\right]^{-1}.
\end{eqnarray}
At low energies $m < E < - E_B$, the running coupling $\lambda(E,E_B) > 0$ is
repulsive, it diverges at $E = - E_B$, and becomes attractive 
(i.e.\ $\lambda(E,E_B) < 0$) at high energies $E > - E_B$. The phase shift 
$\delta(k)$ is illustrated in Figure 9. goes through a resonance at $E = - E_B$
with $\delta(\sqrt{E_B^2 - m^2}) = \frac{\pi}{2}$. Since we still have
$\delta(0) = \frac{\pi}{2}$, this behavior is still consistent with the 1-d 
version of Levinson's theorem.
\begin{figure}[t]
\begin{center}
\epsfig{file=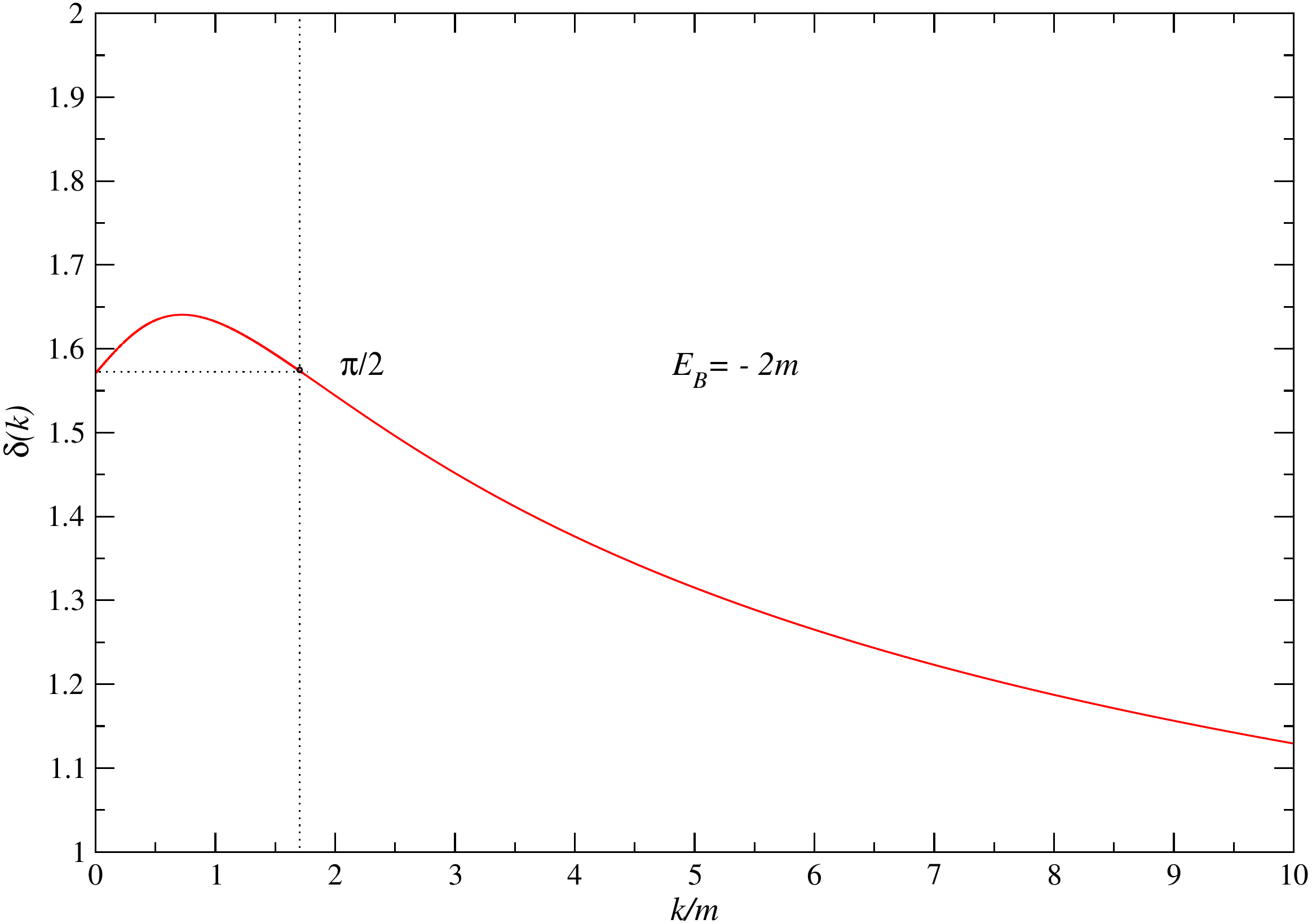,width=0.65\textwidth}
\end{center}
\caption{\it Phase shift $\delta(k)$ as a function of the wave number $k$ for
$E_B = - 2m$. The phase shift goes through a resonance at $E = - E_B$ with
$\delta(k) = \delta(\sqrt{E_B^2 - m^2}) = \frac{\pi}{2}$.}
\end{figure}
\begin{figure}[b]
\begin{center}
\epsfig{file=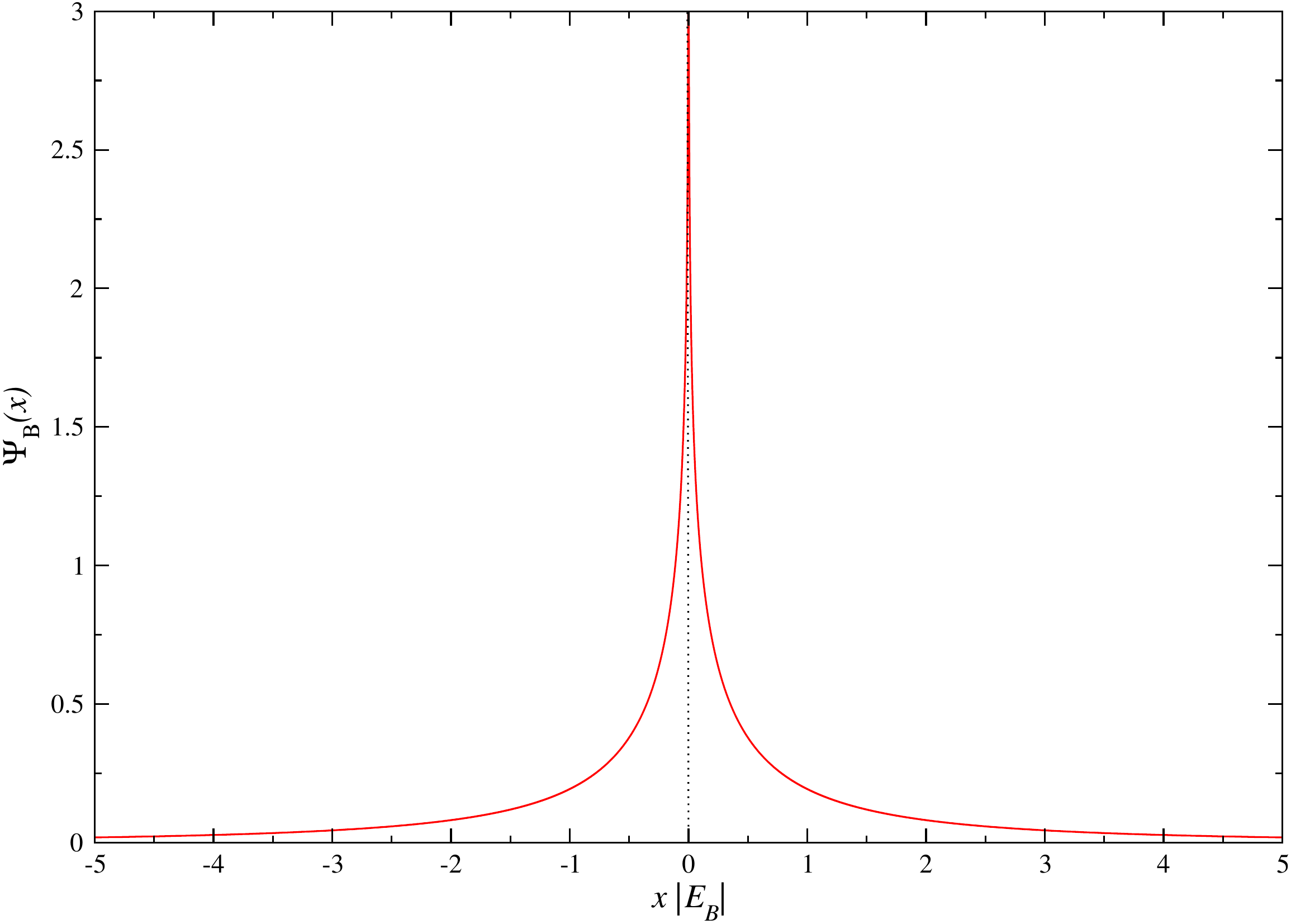,width=0.65\textwidth}
\end{center}
\caption{\it Bound state wave function in the massless case.}
\end{figure}

\section{The Massless Case}

Let us also consider the massless case $m = 0$. Since $\lambda$ is 
dimensionless, the system is then scale-invariant, at least at the classical 
level. For $m = 0$, we are automatically limited to ultra-strong bound states 
(with $E_B < - m$). The bound state energy $E_B$ is a scale generated
non-perturbatively at the quantum level, in a similar way as the proton mass
is generated in massless QCD. Scale invariance is then anomalously broken and
a scale, in this case $E_B$, emerges by dimensional transmutation. All physical
quantities can then be expressed in units of this scale.

First, let us consider the bound state wave function in momentum space
\begin{equation}
\widetilde \Psi_B(p) = \sqrt{\frac{\pi}{- E_B}} \ \frac{1}{E_B - |p|}.
\end{equation}
In coordinate space, it takes the form
\begin{equation}
\Psi_B(x) = \sqrt{\frac{1}{- \pi E_B}} 
\int_0^\infty d\mu \ \frac{\mu}{\mu^2 + E_B^2} \exp(- \mu |x|),
\end{equation}
which is illustrated in Figure 10.
\begin{figure}[t]
\begin{center}
\epsfig{file=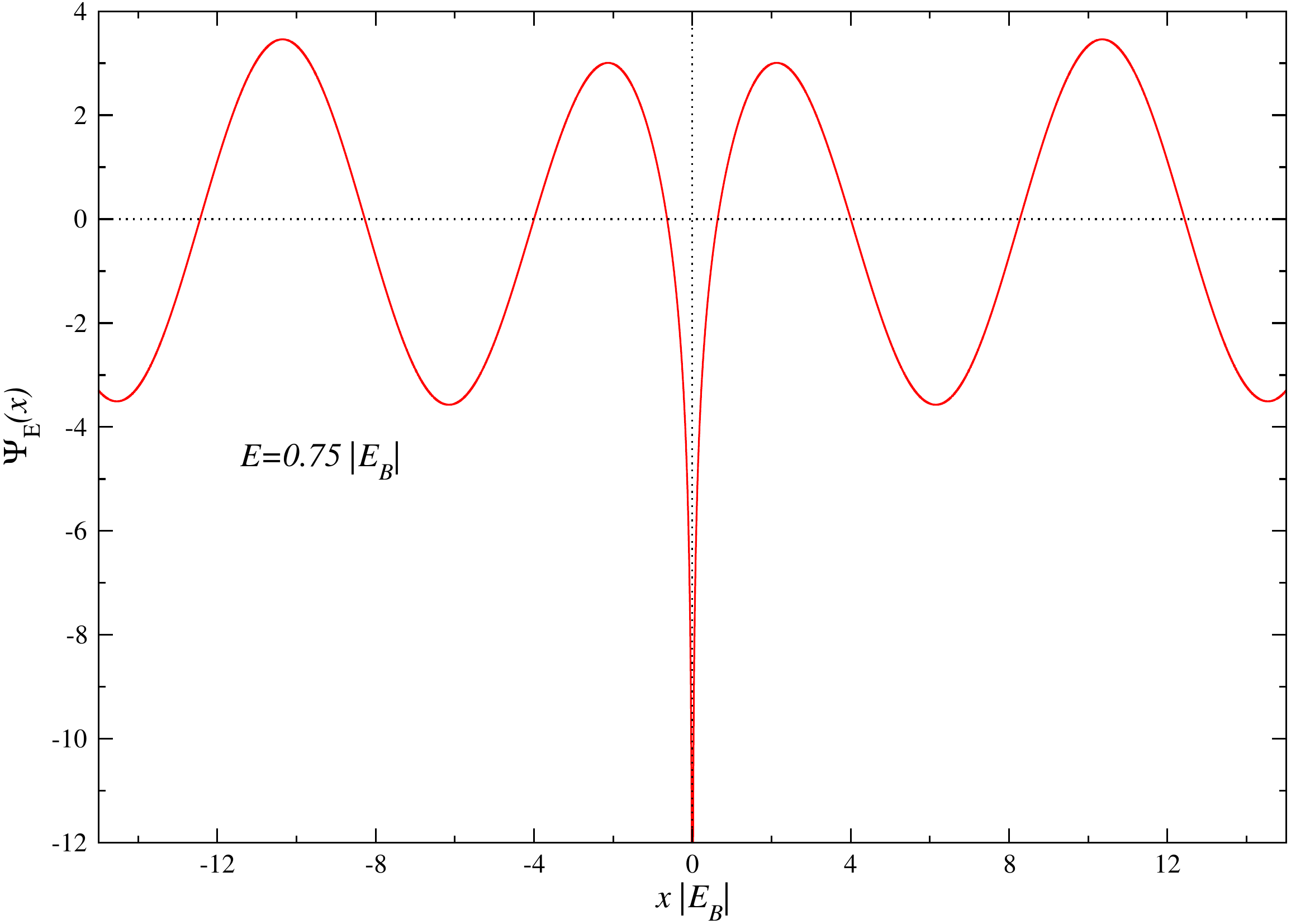,width=0.65\textwidth} \vskip0.5cm
\epsfig{file=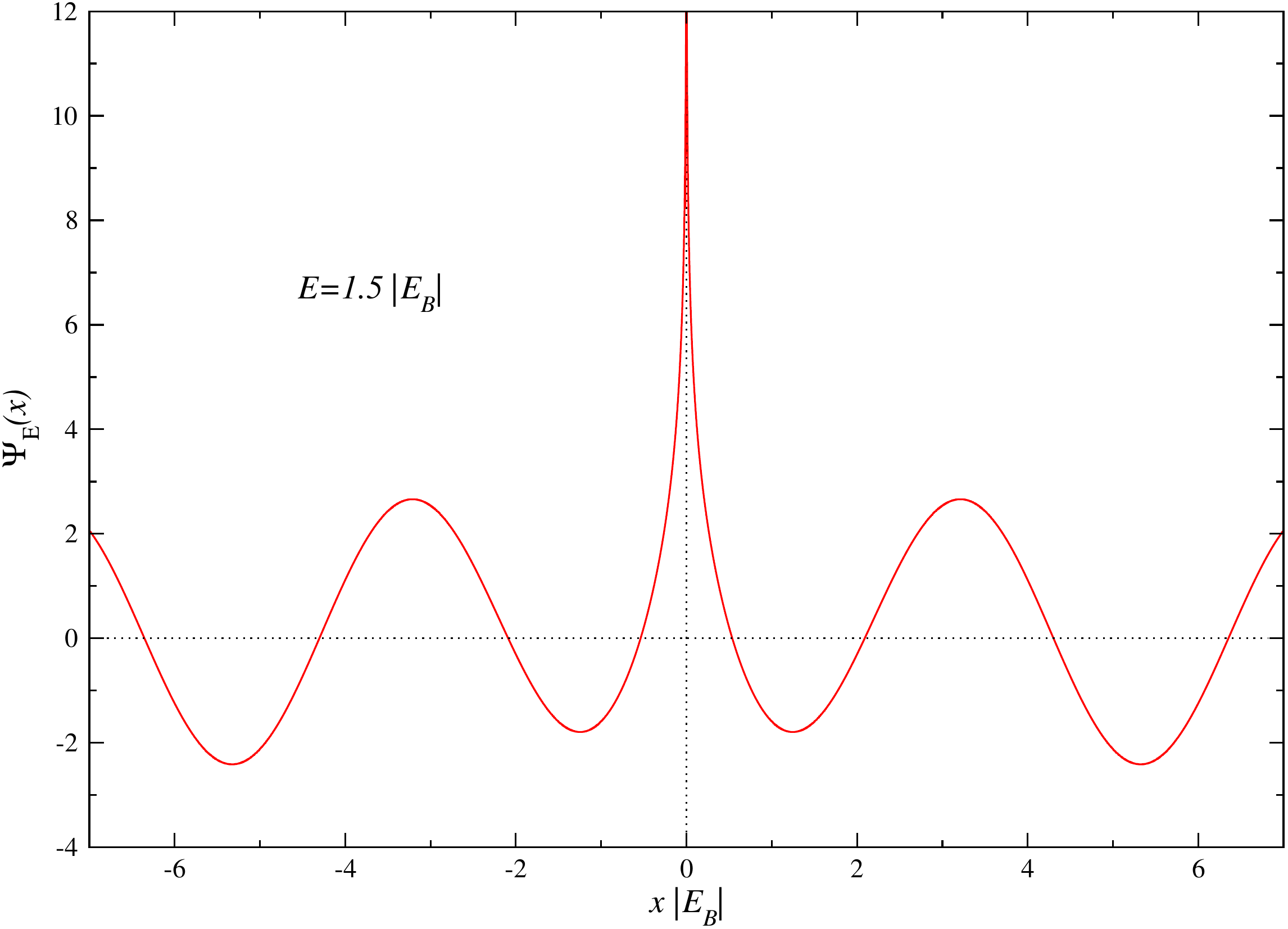,width=0.65\textwidth}
\end{center}
\caption{\it Scattering wave functions in the massless case, for 
$E = 0.75 |E_B| < - E_B$ (top panel), and for $E = 1.5 |E_B| > - E_B$ (bottom
panel).}
\end{figure}
As usual, the wave function diverges logarithmically at $x = 0$, but is still 
square-integrable. Next, we consider the even-parity scattering state (with
$E = k$)
\begin{eqnarray}
\Psi_E(x)&=&A(k) \left[ \cos(k x) + \lambda(E,E_B) \sin(k |x|) \right. 
\nonumber \\
&-&\left. \frac{\lambda(E,E_B)}{\pi} \int_0^\infty d\mu 
\frac{\mu}{\mu^2 + k^2} \exp(- \mu |x|) \right].
\end{eqnarray}
Two scattering wave functions, one for $m < E < - E_B$ and one for $E > - E_B$, 
are shown in Figure 11.

The resulting reflection and transmission amplitudes as well as the S-matrix 
are then given by
\begin{equation}
R(k) = - \frac{i \lambda(E,E_B)}{1 + i \lambda(E,E_B)}, \quad
T(k) = \frac{1}{1 + i \lambda(E,E_B)}, \quad
S(k) = \frac{1 - i \lambda(E,E_B)}{1 + i \lambda(E,E_B)}.
\end{equation}
In the massless case, one obtains
\begin{equation}
\tan\delta(k) = - \lambda(E,E_B) = \frac{\pi}{\log(- E/E_B)} 
= \frac{\pi}{\log(k/|E_B|)}.
\end{equation}
The $\beta$-function then reduces to
\begin{equation}
\label{betamassless}
\beta(\lambda) = E \frac{\partial |\lambda(E,E_B)|}{\partial E} =
- \frac{\pi}{(\log(- E/E_B))^2} = - \frac{\lambda(E,E_B)^2}{\pi},
\end{equation}
which is now valid even at low energies. The running coupling and the 
$\beta$-function are shown in Figure 12. Remarkably, the running coupling 
vanishes not only at high, but also at low energies. In fact, the theory has
both an ultra-violet and an infra-red fixed point. At the ultra-violet fixed
point, $\lambda(E,E_B)$ approaches 0 from below, as $E \rightarrow \infty$,
while at the infra-red fixed point, $\lambda(E,E_B)$ approaches 0 from above, as
$E \rightarrow 0$. Both fixed points are described by the same zero of the
$\beta$-function of eq.(\ref{betamassless}).
\begin{figure}[t]
\begin{center}
\epsfig{file=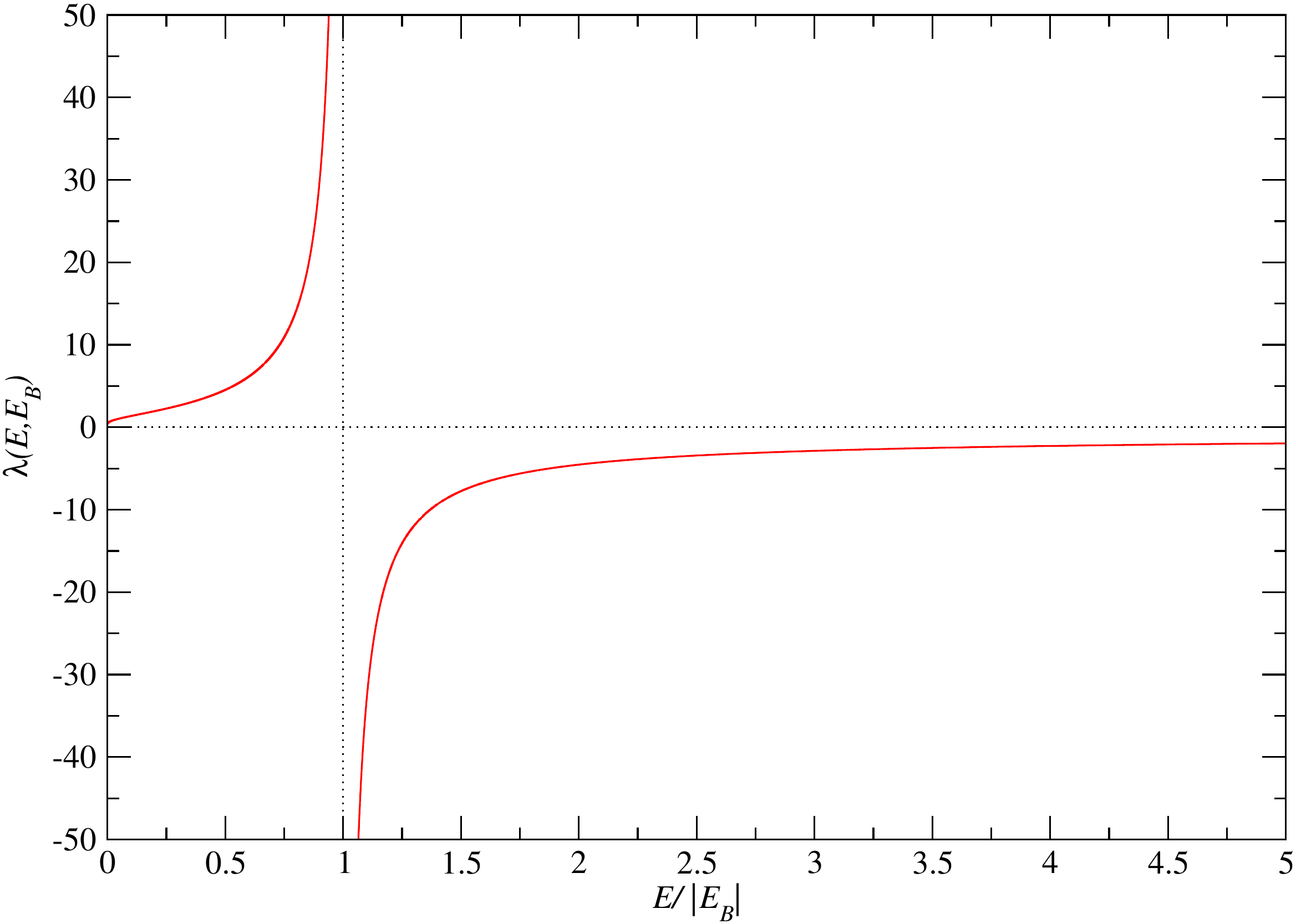,width=0.65\textwidth} \vskip0.5cm
\epsfig{file=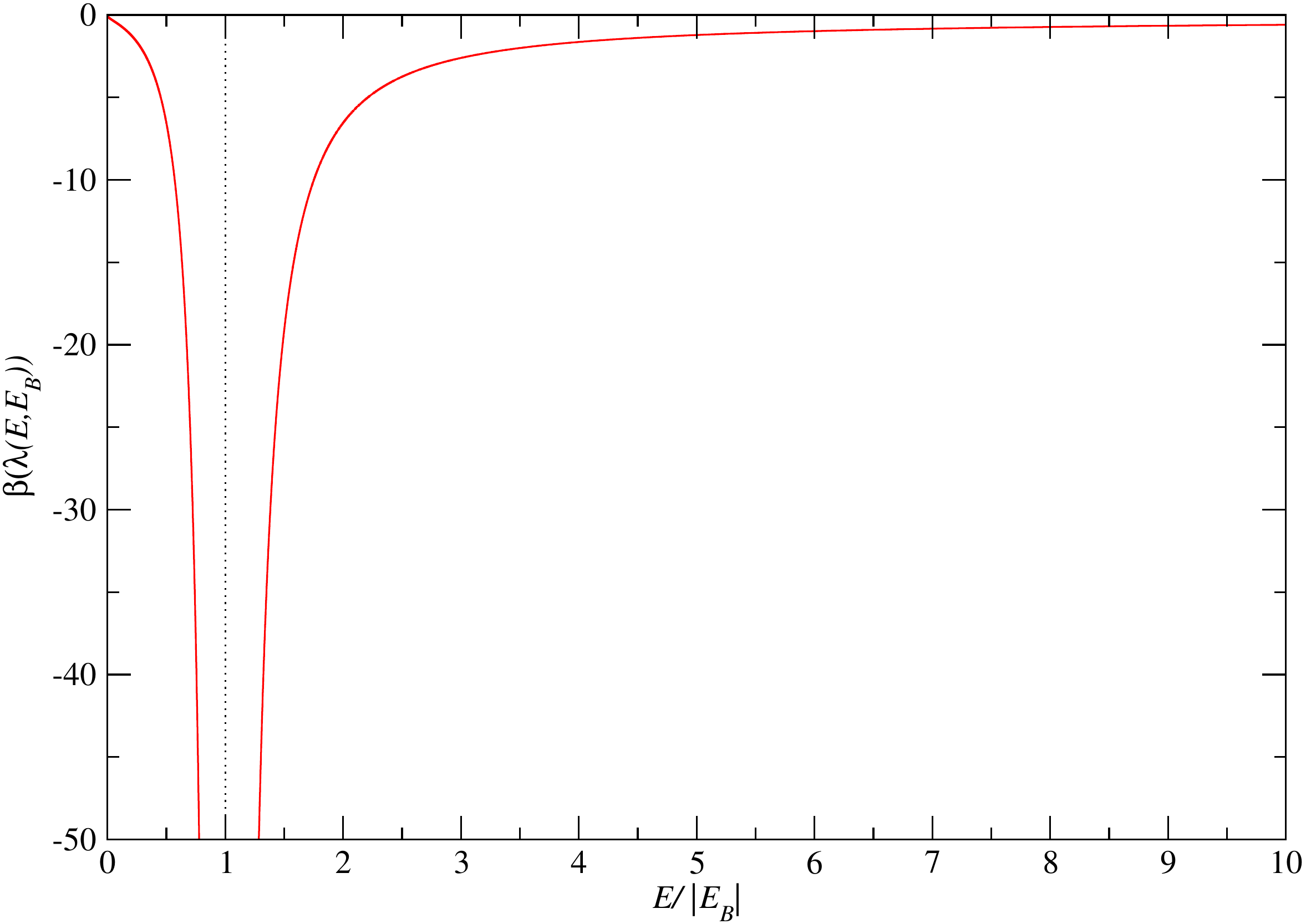,width=0.65\textwidth}
\end{center}
\caption{\it Top panel: The running coupling $\lambda(E,E_B)$ as a function of
the energy $E$ in the massless case. The coupling goes to zero both at high and
at low energies. Bottom panel: The $\beta$-function $\beta(\lambda(E,E_B))$ as 
a function of the scattering energy $E$ (in units of $|E_B|$) in the massless 
limit.}
\end{figure}

This situation resembles the one of an asymptotically free non-Abelian gauge
theory near the so-called conformal window, which is relevant in the context of 
walking technicolor theories \cite{Hol81,Ban82,App87,Die05,Gie06,Kap09}.
Another system of this kind is the 2-d $O(3)$ model at vacuum angle 
$\theta = \pi$ \cite{Zam86,deF12}, whose low-energy effective theory is the 
conformal $k=1$ Wess-Zumino-Novikov-Witten model \cite{Wes71,Nov81,Wit84}. Such 
theories also have both an ultra-violet and an infra-red fixed point. While the 
theory is scale invariant at very low energies, scale invariance is still 
explicitly violated, via dimensional transmutation, at a non-perturbatively 
generated higher energy scale. Thanks to asymptotic freedom, this scale is 
exponentially small compared to the ultimate ultra-violet cut-off (which can 
thus be sent to infinity). In our model, the energy $E_B < 0$ of the bound 
state sets the non-perturbatively generated energy scale, which still affects 
the scattering states at high energies $E > - E_B$. Low-energy scattering states
(with $0 < E \ll - E_B$) are governed by the infra-red fixed point and can thus 
be mapped into each other by scale transformations, as illustrated in Figure 13.
\begin{figure}[t]
\begin{center}
\epsfig{file=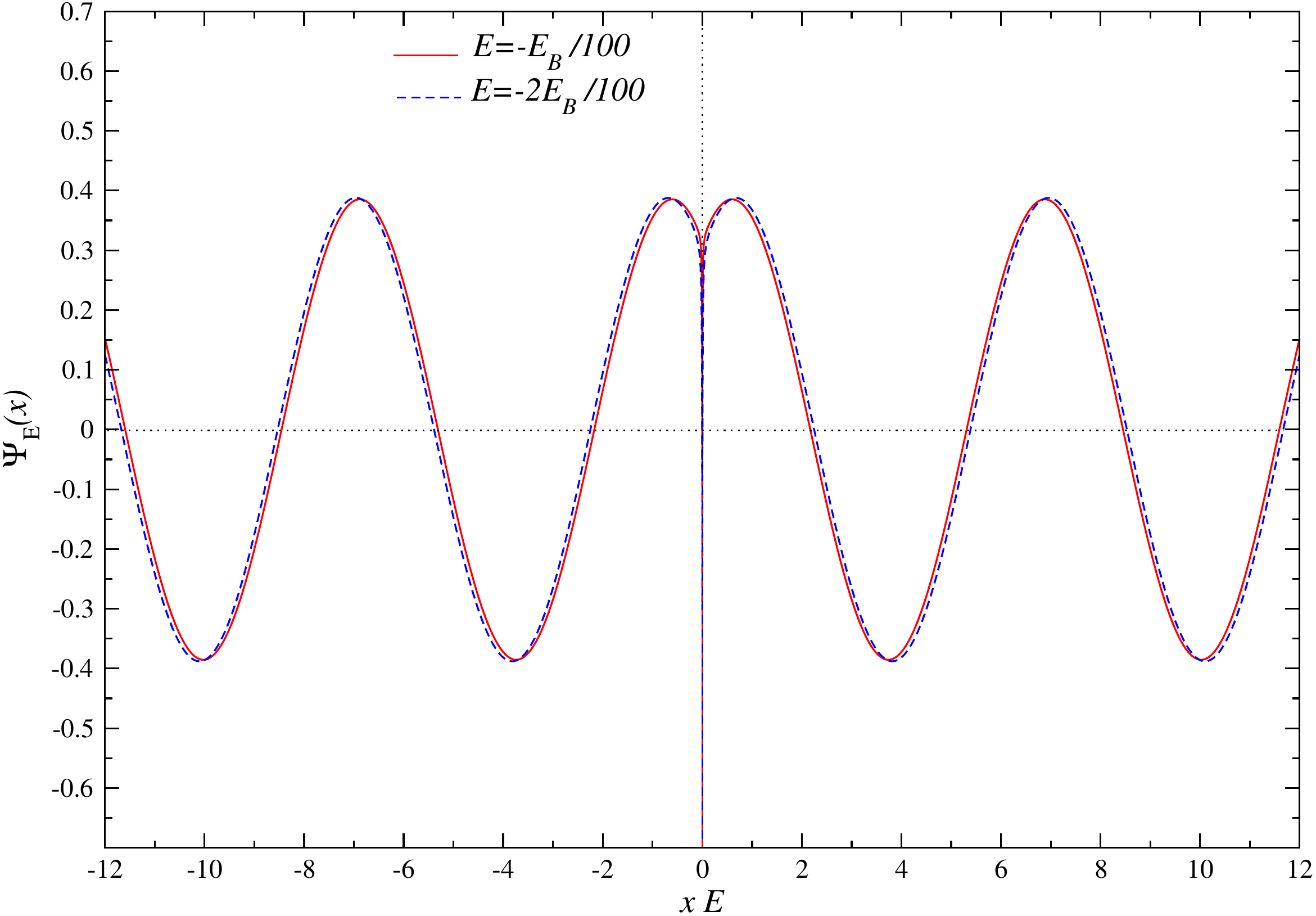,width=0.65\textwidth} \vskip0.5cm
\epsfig{file=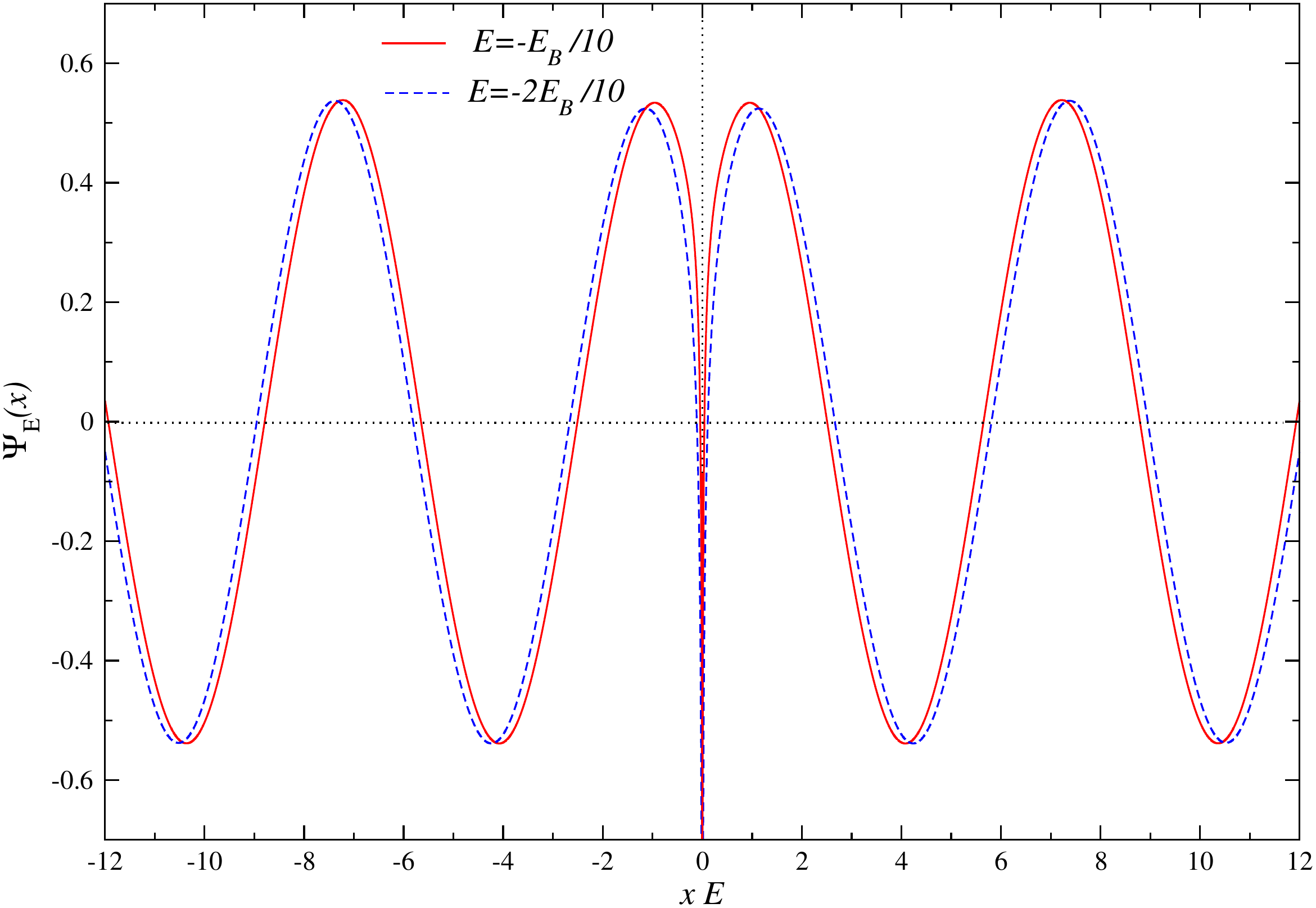,width=0.65\textwidth}
\end{center}
\caption{\it Top panel: Even-parity scattering wave functions at low energies 
$E = |E_B|/100$ and $E = 2 |E_B|/100$, very close to the infra-red conformal 
fixed point, as a function of the rescaled position $x E$. The two wave 
functions are related by a factor 2 scale transformation. Bottom panel: 
The scattering wave functions at somewhat higher energies $E = |E_B|/10$ and 
$E = 2 |E_B|/10$, further away from the conformal fixed point, show a visible 
deviation from scale invariance.}
\end{figure}

\section{Conclusions}

We have investigated contact interactions in 1-dimensional relativistic quantum
mechanics. In contrast to the non-relativistic case, there is only a 1-parameter
family of self-adjoint extensions of the pseudo-differential operator 
$H = \sqrt{p^2 + m^2}$, which is characterized by the contact potential 
$\lambda \delta(x)$. Remarkably, this simple potential gives rise to rather
rich physics. First of all, unlike in the non-relativistic case, the 
$\delta$-function potential requires regularization and subsequent 
renormalization, which we have performed using dimensional regularization. 
Indeed, using this physics approach, we obtained results that are consistent 
with the more abstract mathematical theory of self-adjoint extensions of 
pseudo-differential operators. That theory also implies that there are no 
non-trivial relativistic contact interactions in more than one spatial 
dimension. This is again in contrast to the non-relativistic case, in which 
there is a 1-parameter family of non-trivial contact interactions both in two 
and three spatial dimensions. In four and more spatial dimensions, on the other
hand, there are no non-trivial self-adjoint extensions of the non-relativistic
free particle Hamiltonian. It is interesting to investigate contact interactions
in higher dimensions also using dimensional regularization. This has already 
been done in the non-relativistic case. While dimensional regularization 
provides results that are consistent with the self-adjoint extension theory in
two and three spatial dimensions, in contrast to the theory of self-adjoint 
extensions, it seems to lead to non-trivial contact interactions in higher 
dimensions \cite{Phi98}. However, it turns out that the resulting Hamiltonian 
is not self-adjoint and thus not physically meaningful. In this sense, 
dimensional regularization actually fails to produce the correct result. We 
suspect that the same may happen in the relativistic case, already in two and 
three spatial dimensions, which might be worth investigating. 

As we discussed before, the external $\delta$-function potential can be
attributed to an infinitely heavy particle. It is interesting to ask whether 
this second particle can be treated fully dynamically, by giving it a finite 
mass. Only then the system may become Poincar\'e invariant, because translation
invariance is no longer explicitly broken by the position of the external
contact interaction. Leutwyler's non-interaction theorem suggests that 
Poincar\'e invariance is incompatible with interacting point particles. However,
since the theorem operates at the classical level, and does not apply to quantum
mechanical point interactions, there may be a quantum loop-hole that would be
worth exploring. For the fully dynamical two-particle problem, the question
arises whether both a self-adjoint Hamiltonian and a self-adjoint boost operator
can be constructed, which obey the commutation relations of the Poincar\'e
algebra together with the operator of the total momentum $P$. If so, the 
two-particle system will have a total energy $E = \sqrt{P^2 + M^2}$, where $M$
is the rest-energy of the system. In such a system, one could also investigate 
the Lorentz contraction of a moving wave packet, which, until now, has been 
investigated for free particles only \cite{AlH09}. Although we know that Nature 
makes relativistic ``particles'' as non-local quantized field excitations, at 
least for pedagogical reasons, it is interesting to explore the alternative 
possibilities of local relativistic point particles. Based on the 
non-interaction theorem, such alternatives are expected to be very limited, 
which, in turn, underscores the strengths of relativistic quantum field 
theories.

As we have shown, asymptotic freedom, dimensional transmutation, and an 
infra-red conformal fixed point in the massless limit, already arise in 
1-dimensional relativistic point particle quantum mechanics with a 
$\delta$-function potential. This allowed us to illustrate non-trivial quantum 
field theoretical phenomena as well as techniques including dimensional 
regularization and renormalization, avoiding the technical complications of 
quantum field theory. We conclude this paper by expressing our hope that the 
relatively simple system that we have investigated here will help to bridge the 
large gap that separates non-relativistic quantum mechanics from relativistic 
quantum field theory in the teaching of fundamental physics.

\section*{Acknowledgments}

This publication was made possible by the NPRP grant \# NPRP 5 - 261-1-054 from
the Qatar National Research Fund (a member of the Qatar Foundation). The
statements made herein are solely the responsibility of the authors.

\end{document}